\documentclass[unnumsec,webpdf,contemporary,large]{oup-authoring-template}

\graphicspath{{Fig/}}

\theoremstyle{thmstyleone}%
\theoremstyle{thmstyletwo}%
\theoremstyle{thmstylethree}%

\usepackage{amsmath}
\usepackage{subfig}
\usepackage{graphicx}
\usepackage[colorinlistoftodos]{todonotes}

\usepackage{bm}
\usepackage{amsfonts} 
\usepackage{amssymb} 
\usepackage{stmaryrd} 

\usepackage{mathtools}

\usepackage{booktabs}
\usepackage{longtable}
\usepackage{array}
\usepackage{multirow}
\renewcommand{\arraystretch}{1}

\usepackage[linesnumbered,ruled,vlined]{algorithm2e}
\makeatletter
\renewcommand{\@algocf@capt@plain}{above}
\makeatother

\usepackage{todonotes}
\usepackage{comment}
 
\usepackage{color}
\definecolor{Orange}{rgb}{1,0.5,0}
\definecolor{Red}{rgb}{1,0,0}
\definecolor{Blue}{rgb}{0,0,1}

\newcommand{\FM}[1]{\textsf{\textbf{\textcolor{magenta}{\footnotesize [FM: #1]}}}}

\begin{document}

\journaltitle{Briefings in Bioinformatics}
\DOI{DOI HERE}
\copyrightyear{2022}
\pubyear{2019}
\access{Advance Access Publication Date: Day Month Year}
\appnotes{Problem Solving Protocol}

\firstpage{1}


\title[Multiple Similarity DTI prediction with RW and MF]{Multiple Similarity Drug-Target Interaction Prediction with Random Walks and Matrix Factorization}

\author[1,2,$\ast$]{Bin~Liu}
\author[2]{Dimitrios~Papadopoulos}
\author[3]{Fragkiskos~D.~Malliaros}
\author[2]{Grigorios~Tsoumakas}
\author[2]{Apostolos~N.~Papadopoulos}

\authormark{Liu et al.}

\address[1]{Key Laboratory of Data Engineering and Visual Computing, Chongqing University of Posts and Telecommunications, Chongqing 400065, China}
\address[2]{School of Informatics, Aristotle University of Thessaloniki, 54124 Thessaloniki, Greece}
\address[3]{Paris-Saclay University, CentraleSupélec, Inria, Centre for Visual Computing (CVN), 91190 Gif-Sur-Yvette, France}

\corresp[$\ast$]{Corresponding author: Bin Liu, E-mail:\href{liubin@cqupt.edu.cn}{liubin@cqupt.edu.cn}}

\received{Date}{0}{Year}
\revised{Date}{0}{Year}
\accepted{Date}{0}{Year}



\abstract{
The discovery of drug-target interactions (DTIs) is a very promising area of research with great potential. The accurate identification of reliable interactions among drugs and proteins via computational methods, which typically leverage heterogeneous information retrieved from diverse data sources, can boost the development of effective pharmaceuticals. Although random walk and matrix factorization techniques are widely used in DTI prediction, they have several limitations. Random walk-based embedding generation is usually conducted in an unsupervised manner, while the linear similarity combination in matrix factorization distorts individual insights offered by different views. To tackle these issues, we take a multi-layered network approach to handle diverse drug and target similarities, and propose a novel optimization framework, called Multiple similarity DeepWalk-based Matrix Factorization (MDMF), for DTI prediction. The framework unifies embedding generation and interaction prediction, learning vector representations of drugs and targets that not only retain higher-order proximity across all hyper-layers and layer-specific local invariance, but also approximate the interactions with their inner product. Furthermore, we develop an ensemble method (MDMF2A) that integrates two instantiations of the MDMF model, optimizing the area under the precision-recall curve (AUPR) and the area under the receiver operating characteristic curve (AUC) respectively. The empirical study on real-world DTI datasets shows that our method achieves statistically significant improvement over current state-of-the-art approaches in four different settings. Moreover, the validation of highly ranked non-interacting pairs also demonstrates the potential of MDMF2A to discover novel DTIs.
}


\maketitle

\section{Introduction}
The main objective of the drug discovery process is to identify drug-target interactions (DTIs) among numerous candidates. Although \textit{in vitro} experimental testing can verify DTIs, it suffers from extremely high time and monetary costs. 
Computational (\textit{in silico}) methods employ machine learning techniques~\cite{Bagherian2021MachinePaper}, such as matrix factorization (MF)~\cite{Ezzat2017Drug-targetFactorization}, kernel-based models~\cite{Ding2020IdentificationFusion}, graph/network embedding~\cite{An2021AInteractions}, and deep learning~\cite{Xuan2021IntegratingPrediction}, to efficiently infer a small amount of candidate drugs. This vastly shrinks the search scope and reduces the workload of experiment-based verification, thereby accelerating the drug discovery process significantly.

In the past, the chemical structure of drugs and the protein sequence of targets were the main source of information for inferring candidate DTIs~\cite{Liu2021Drug-targetRecovery,Liu2016NeighborhoodPrediction,Pliakos2021PredictingPartitioning}. 
Recently, with the advancements in clinical medical technology, abundant drug and target-related biological data from multifaceted sources are exploited to boost the accuracy of DTI prediction. Some MF and kernel-based methods utilize multiple types of drug and target similarities derived from heterogeneous information by integrating them into a single drug and target similarity~\cite{Zheng2013CollaborativeInteractions, Ding2020IdentificationFusion, liu2021optimizing, Olayan2018DDR:Approaches},
but in doing so discard the distinctive information possessed by each similarity view. 

In contrast, network-based approaches consider the diverse drug and target data as a (multiplex) heterogeneous DTI network that describes multiple aspects of drug and target relations, and learn topology-preserving representations of drugs and targets to facilitate DTI prediction. 
With deep neural networks showing consistently superior performance in the latest years in a plethora of different learning tasks, their adoption in the DTI prediction field, especially inferring new DTIs by mining DTI networks, is understandably rising~\cite{Wan2019NeoDTI:Interactions, Xuan2021IntegratingPrediction, chen2022SupDTI, chen2022DCFME}. 
Although deep learning models achieve improved performance, they require larger amounts of data and are computationally intensive~\cite{Bagherian2021MachinePaper}. In addition, most deep learning models, are sensitive to noise~\cite{dai_adversarial_2018}. This is very important in DTI prediction, since there are many undiscovered interactions in the  bipartite network of drugs and targets~\cite{Liu2021Drug-targetRecovery,Pliakos2020Drug-targetReconstruction,Ezzat2017Drug-targetFactorization}.


Apart from deep learning, another type of network-based model widely used in DTI prediction computes graph embeddings based on {\em random walks}~\cite{Luo2017AInformation,An2021AInteractions}.
Although these methods can model high-order node proximity efficiently, they typically perform embedding generation and interaction prediction as two independent tasks. Hence, their embeddings are learned in an unsupervised manner, failing to preserve the topology information from the interaction network. 

Random walk embedding methods are essentially factorizing a matrix capturing node co-occurrences within random walk sequences generated from the graph~\cite{qiu2018network}---allowing to unify embedding generation and interaction prediction under a common MF framework. 
Nevertheless, the MF method proposed in~\cite{qiu2018network} that approximates DeepWalk~\cite{Perozzi2014DeepWalk:Representations} can only handle single-layer networks. Thus, it is unable to fully exploit the topology information of multiple drug and target layers present in multiplex heterogeneous DTI networks.
Furthermore, the area under the precision-recall curve (AUPR) and the area under the receiver operating characteristic curve (AUC) are two important evaluation metrics in DTI prediction, but no network-based approach directly optimizes them.


To address the issues mentioned above, we propose the formulation of a DeepWalk-based MF model, called Multiple similarity DeepWalk-based Matrix Factorization (MDMF), which incorporates multiplex heterogeneous DTI network embedding generation and DTI prediction within a unified optimization framework. It learns vector representations of drugs and targets that not only capture the multilayer network topology via factorizing the hyper-layer DeepWalk matrix with information from diverse data sources, but also preserve the layer-specific local invariance with the graph Laplacian for each drug and target similarity view. In addition, the DeepWalk matrix contains richer interaction information, exploiting high-order node proximity and implicitly recovering the possible missing interactions. Based on this formulation, we instantiate two models that leverage surrogate losses to optimize two essential evaluation measures in DTI prediction, namely AUPR and AUC. In addition, we integrate the two models to consider the maximization of both metrics. Experimental results on DTI datasets under various prediction settings show that the proposed method outperforms state-of-the-art approaches and can discover new reliable DTIs. 

The rest of this article is organized as follows.  Section~\ref{sec.preliminaries} introduces some preliminaries of our work. Section~\ref{sec.proposed} presents the proposed approach. Performance evaluation results and relevant discussions are offered in Section~\ref{sec.experiments}. Finally, Section~\ref{sec.conclusions} concludes this work.


\section{Preliminaries} 
\label{sec.preliminaries}

\subsection{Problem Formulation}
Given a drug set $D=\{d_i\}_{i=1}^{n_d}$ and a target set $T=\{t_i\}_{i=1}^{n_t}$, the relation between drugs (targets) can be assessed in various aspects, which are represented by a set of similarity matrices $\{\bm{S}^{d,h}\}_{h=1}^{m_d}$ ($\{\bm{S}^{t,h}\}_{h=1}^{m_t}$), where $\bm{S}^{d,h} \in \mathbb{R}^{n_d \times n_d}$ ($\bm{S}^{t,h} \in \mathbb{R}^{n_t \times n_t}$) and $m_d$ ($m_t$) is the number of relation types for drugs (targets). In addition, let the binary matrix $\bm{Y} \in \{0,1\}^{n_d \times n_t}$ indicate the interactions between drugs in $D$ and targets in $T$, where $Y_{ij}=1$ denotes that $d_i$ and $t_j$ interact with each other, and $Y_{ij} = 0$ otherwise.
A DTI dataset for $D$ and $T$ consists of $\{\bm{S}^{d,h}\}_{h=1}^{m_d}$, $\{\bm{S}^{t,h}\}_{h=1}^{m_t}$ and $\bm{Y}$.

Let ($d_x$,$t_z$) be a test drug-target pair, $\{\bm{\bar{s}}^{d,h}_x\}^{m_d}_{h=1}$ be a set of $n_d$-dimensional vectors storing the similarities between $d_x$ and $D$, and $\{\bm{\bar{s}}^{t,h}_z\}^{m_d}_{h=1}$ be a set of $n_t$-dimensional vectors storing the similarities between $t_z$ and $T$. A DTI prediction model predicts a real-valued score $\hat{Y}_{xz}$ indicating the confidence of the affinity between $d_x$ and $t_z$. In addition, $d_x \notin D$ ($t_z \notin T$), which does not belong to the training set, is considered as the new drug (target).
There are four prediction settings according to whether the drug and target involved in the test pair are training entities~\cite{Pahikkala2015TowardPredictions}: 
\begin{itemize}
    \itemsep-0.1em 
    \item S1: predict the interaction between $d_{x} \in D$ and $t_z \in T$;
    \item S2: predict the interaction between $d_{x} \notin D$ and $t_z \in T$;
    \item S3: predict the interaction between $d_x \in D$ and $t_z \notin T$;
    \item S4: predict the interaction between $d_x \notin D$ and $t_z \notin T$.
\end{itemize}

\subsection{Matrix Factorization for DTI Prediction} 
\label{sec:MFDTI} 
In DTI prediction, MF methods typically learn two vectorized representations of drugs and targets that approximate the interaction matrix $\bm{Y}$ by minimizing the following objective:
\begin{equation}
    \min_{\bm{U},\bm{V}} \mathcal{L}(\hat{\bm{Y}},\bm{Y}) + \mathcal{R}(\bm{U},\bm{V}),
    \label{eq:MF_obj}
\end{equation}
where $\hat{\bm{Y}} = f(\bm{U}\bm{V}^\top) \in \mathbb{R}^{n_d \times n_t}$ is the predicted interaction matrix, $f$ is either the identity function $\omega$ for standard MF~\cite{Ezzat2017Drug-targetFactorization} or the element-wise logistic function $\sigma$ for Logistic MF~\cite{Liu2016NeighborhoodPrediction}, and $\bm{U} \in \mathbb{R}^{n_d \times r}$, $\bm{V} \in \mathbb{R}^{n_t \times r}$ are $r$-dimensional drug and target latent features (embeddings), respectively.
The objective in Eq. \eqref{eq:MF_obj} includes two parts: $\mathcal{L}(\hat{\bm{Y}},\bm{Y})$ is the loss function to evaluate the inconsistency between the predicted and ground truth interaction matrix, and $\mathcal{R}(\bm{U},\bm{V})$ concerns the regularization of the learned embeddings.

Given a test drug-target pair ($d_x,t_z$), its prediction with a specific instantiation of $f$ is computed based on the embeddings of $d_x$ ($\bm{U}_x \in \mathbb{R}^{r}$) and $t_z$ ($\bm{V}_z \in \mathbb{R}^{r}$):
\begin{equation}
\hat{Y}_{xz} = \left\{ 
\begin{aligned}
&\bm{U}_x\bm{V}_z^\top, \text{if } f = \omega \\
&\left(1+\exp(-\bm{U}_x\bm{V}_z^\top)\right)^{-1}, \text{if } f = \sigma.
\end{aligned}
\right. 
\end{equation}


\subsection{DeepWalk Embeddings as Matrix Factorization} \label{sec:DW}
DeepWalk~\cite{Perozzi2014DeepWalk:Representations} is a network embedding approach, which generates a number of random walks over a graph to capture proximity among nodes. 
Qiu et al. \cite{qiu2018network} proved that the DeepWalk model could be interpreted as a matrix factorization task when the length of random walk approaches infinity. In particular, they introduced NetMF, a model that approximates DeepWalk to learn embeddings of a network $G$ containing $n$ nodes by factorizing the DeepWalk matrix defined as:
\begin{equation}
    \bm{M} =\log  \left( \max \left(1, \frac{\psi(\bm{A})}{n_sn_w} \left(\sum_{h=1}^{n_w} (\bm{\Lambda}^{-1} \bm{A})^h \right) \bm{\Lambda}^{-1} \right) \right),
    \label{eq:M}
\end{equation}
where $\bm{A} \in \mathbb{R}^{n \times n}$ is the adjacency matrix of $G$, $\psi(\bm{A}) = \sum_i\sum_j A_{ij}$, $\bm{\Lambda}= \text{diag}(\bm{Ae})$ is a diagonal matrix with row sum of $\bm{A}$, $n_w$ is the window size of the random walk controlling the number of context nodes, $n_s$ plays the same role as the number of negative samples in DeepWalk, while the $\max$ function guarantees that all elements in $\bm{M}$ being non-negative.
Considering the symmetry of $\bm{M}$ for undirected networks, the factorization of the DeepWalk matrix could be expressed as $\bm{M}=\bm{Q}\bm{Q}^\top$, where $\bm{Q} \in \mathbb{R}^{n \times r}$ represents the $r$-dimensional network embeddings.

\section{Materials and Methods}
\label{sec.proposed}


\subsection{Datasets}
Two types of DTI datasets, constructed based on online biological and pharmaceutical databases, are used in this study. Their characteristics are shown in Table \ref{tab:Dataset}. 

\begin{table}[h]
\centering
\caption{Characteristics of datasets.}
\label{tab:Dataset}
\begin{tabular*}{\columnwidth}{@{\extracolsep{\fill}}ccccccc@{\extracolsep{\fill}}}
\toprule
Dataset & $n_d$ & $n_t$ & $|P_1|$ & Sparsity & $m_d$ & $m_t$ \\ \midrule
\textsl{NR} & 54 & 26 & 166 & 0.118 & \multirow{4}{*}{4} & \multirow{4}{*}{4} \\
\textsl{GPCR} & 223 & 95 & 1096 & 0.052 &  &  \\
\textsl{IC} & 210 & 204 & 2331 & 0.054 &  &  \\
\textsl{E} & 445 & 664 & 4256 & 0.014 &  &  \\ \midrule
\textsl{Luo} & 708 & 1512 & 1923 & 0.002 & 4 & 3 \\ \bottomrule
\end{tabular*}
\end{table}

The first one is a collection of four golden standard datasets constructed by Yamanishi et al. ~\cite{Yamanishi2008PredictionSpaces}, each one corresponding to a target protein family, namely Nuclear Receptors (\textsl{NR}), Ion Channel (\textsl{IC}), G-protein coupled receptors (\textsl{GPCR}), and Enzyme (\textsl{E}).  
Because the interactions in these datasets were discovered 14 years ago, we updated them by adding newly discovered interactions between drugs and targets in these datasets recorded in the last version of KEGG~\cite{Kanehisa2017KEGG:Drugs}, DrugBank~\cite{Wishart2018DrugBank2018}, and ChEMBL~\cite{Mendez2019ChEMBL:Data} databases. Details on new DTIs collection can be found in Supplementary Section A1.
Four types of drug similarities, including SIMCOMP~\cite{Hattori2003DevelopmentPathways} built upon chemical structures,
AERS-freq, AERS-bit~\cite{Takarabe2012DrugApproach} and SIDER~\cite{Kuhn2016TheEffects} derived from drug side effects, as well as four types of target similarities, namely gene ontology (GO) term based semantic similarity, Normalized Smith–Waterman (SW), spectrum kernel with 3-mers length (SPEC-k3), and 4-mers length (SPEC-k4) based amino acid sequence similarities, obtained from~\cite{Nascimento2016APrediction} are utilized to describe diverse drug and target relations, since they possess higher local interaction consistency~\cite{liu2021optimizing}.

The second one provided by Luo et al. \cite{Luo2017AInformation} (denoted as \textsl{Luo}), was built in 2017, includes DTIs and drug-drug interactions (DDI) obtained from DrugBank 3.0~\cite{Wishart2018DrugBank2018}, as well as drug side effect (SE) associations, protein–protein interactions (PPI) and disease-related associations extracted from SIDER~\cite{Kuhn2016TheEffects}, HPRD~\cite{KeshavaPrasad2009HumanUpdate}, and Comparative Toxicogenomics Database~\cite{Davis2013The2013}, respectively. 
Based on diverse drug and target interaction/association profiles, three drug similarities derived from DDI, SE, and drug-disease associations as well as two target similarities derived from PPI and target-disease associations are computed. The Jaccard similarity coefficient is employed to assess the similarity of drugs, SEs or diseases (proteins or diseases) associated/interacted with two drugs (targets). In addition, drug similarity based on chemical structure and target similarity based on genome sequence are also computed. Therefore, four drug similarities and three target similarities are used for this dataset.

\subsection{Multiple Similarity DeepWalk-based Matrix Factorization}

A DTI dataset, associated with multiple drug and target similarities, can be viewed as a multiplex heterogeneous network $G^{DTI}$. This can be done by treating drugs and targets as two types of vertices, and by considering non-zero similarities and interactions as edges connecting two homogeneous and heterogeneous entities respectively, where the weight of each edge equals the corresponding similarity or interaction value.
In a DTI dataset, the interaction matrix is typically more sparse than similarity matrices, causing that similarity derived edges linking two drugs (targets) markedly outnumber more crucial bipartite interaction edges. 
To balance the distribution of different types of edges and stress relations of more similar entities in the DTI network, we replace each original dense similarity matrix with the sparse adjacency matrix of its corresponding  $k$-nearest neighbors ($k$-NNs) graph. Specifically, given a similarity matrix $\bm{S}^{d,h}$, its sparsified matrix $\hat{\bm{S}}^{d,h} \in \mathbb{R}^{n_d \times n_d}$ is defined as:
\begin{equation}
   \hat{S}^{d,h}_{ij} = \left\{
\begin{aligned}
 & S^{d,h}_{ij} \text{, if } d_j \in \mathcal{N}^{k,h}_{d_i} \text{ and } d_i \in \mathcal{N}^{k,h}_{d_j} \\
 & 0 \text{, if } d_j \notin \mathcal{N}^{k,h}_{d_i} \text{ and } d_i \notin \mathcal{N}^{k,h}_{d_j} \\
 & \frac{1}{2}S^{d,h}_{ij} \text{, otherwise,}
\end{aligned}
\right. 
\end{equation}
where $\mathcal{N}^{k,h}_{d_i}$ is the set of the $k$-NNs of $d_i$ based on similarity $\bm{S}^{d,h}$.



\begin{figure*}[h]
\centering
\subfloat[A multiplex heterogeneous DTI network]{\includegraphics[width=0.45\textwidth]{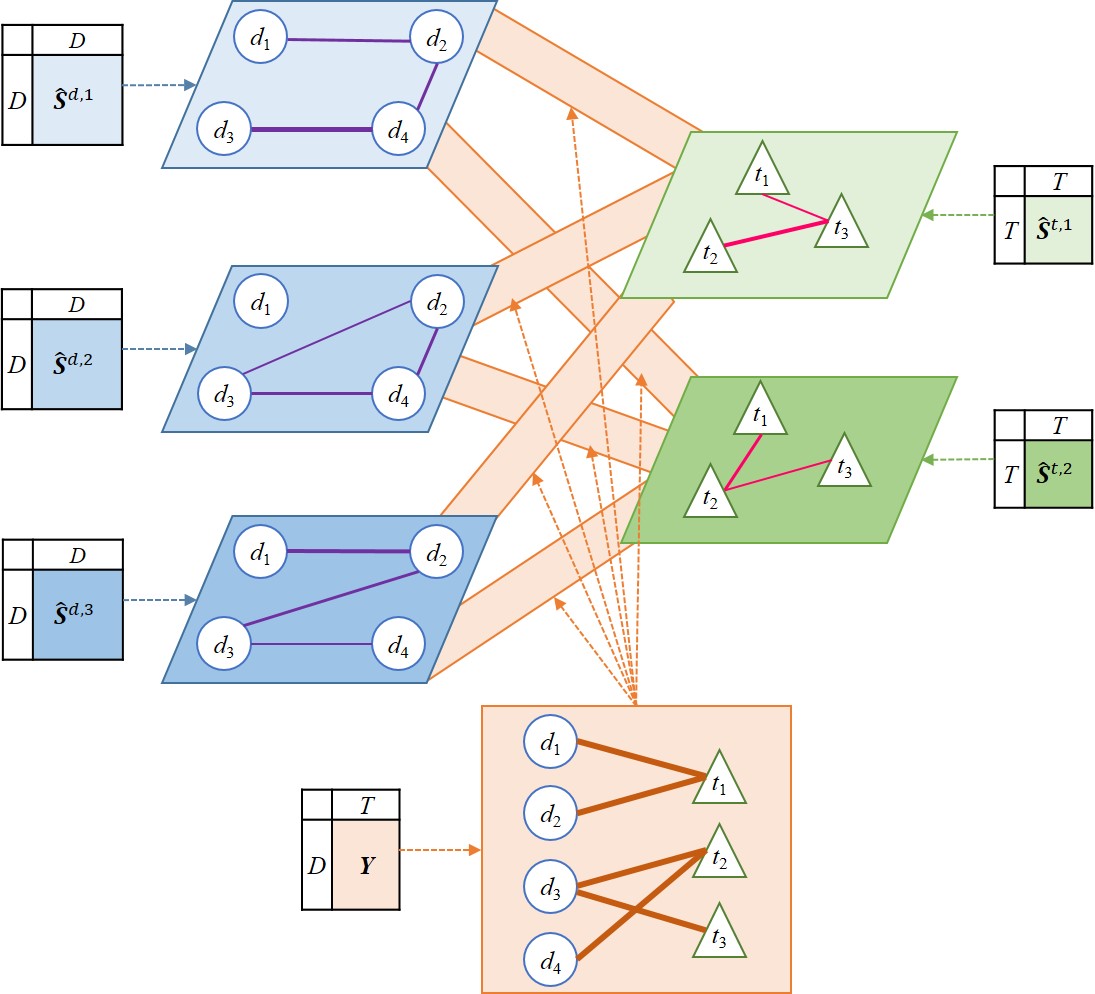}
\label{fig:DTI_network Multipelx_heter}}
\hspace{1em}
\subfloat[The left DTI network reformulated as multiple hyper-layers]{\includegraphics[width=0.35\textwidth]{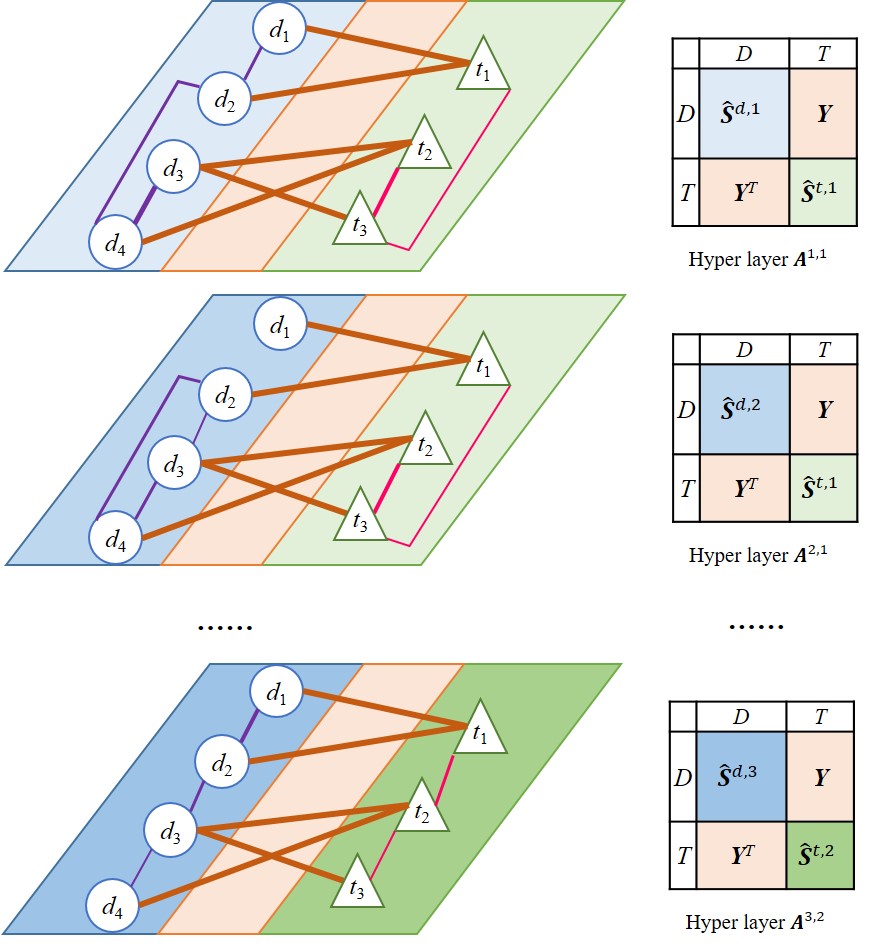}
\label{fig:DTI_network Multi_HyperLayer}}
\caption{Representing a DTI dataset with three drug and two target similarities as a network. (a) A multiplex heterogeneous network including three drug layers, two target layers, and six identical bipartite interaction subnetworks connecting drug and target nodes in each layer. (b) Six multiple hyper-layers, where each of them is composed of a drug and a target layer along with the interaction subnetwork.} 
\label{fig:DTI_network}
\end{figure*}

Formally, $G^{DTI}$ consists of three parts: (i) $G^{d}=\{\hat{\bm{S}}^{d,h}\}_{h=1}^{m_d}$, which is a multiplex drug subnetwork containing $m_d$ layers, with $\hat{\bm{S}}^{d,h}$ being the adjacency matrix of the $h$-th drug layer; (ii) $G^{t}=\{\hat{\bm{S}}^{t,h}\}_{h=1}^{m_t}$, which is a multiplex target subnetwork including $m_t$ layers with $\hat{\bm{S}}^{t,h}$ denoting the adjacency matrix of the $h$-th target layer; (iii) $G^Y=\bm{Y}$, which is a bipartite interaction subnetwork connecting drug and target nodes in each layer. Figure~\ref{fig:DTI_network Multipelx_heter} depicts an example DTI network.  

The DeepWalk matrix cannot be directly calculated for the complex DTI network that includes two multiplex and a bipartite subnetwork. To facilitate its computation, we consider each combination of a drug and a target layer along with the interaction subnetwork as a hyper-layer, and reformulate the DTI network as a multiplex network containing $m_d\cdot m_t$ hyper-layers. 
The hyper-layer incorporating the $i$-th drug layer and $j$-th target layer, is defined  by the adjacency matrix $\bm{A}^{i,j} = \begin{bmatrix} 
\hat{\bm{S}}^{d,i} & \bm{Y} \\
\bm{Y}^\top & \hat{\bm{S}}^{t,j}  \\
\end{bmatrix} $, upon which $G^{DTI}$ could be expressed as a set of hyper-layers $\{\bm{A}^{i,j}\}^{m_d,m_t}_{i=1,j=1}$.  
Figure~\ref{fig:DTI_network Multi_HyperLayer} illustrates the multiple hyper-layer network corresponding to the original DTI network in Figure~\ref{fig:DTI_network Multipelx_heter}.

Based on the above reformulation, we compute a DeepWalk matrix $\bm{M}^{i,j} \in \mathbb{R}^{(n_d+n_t) \times (n_d+n_t)}$ for each $\bm{A}^{i,j}$ using Eq.~\eqref{eq:M}, which reflects node co-occurrences in truncated random walks and captures richer proximity among nodes than the original hyper-layer---especially the proximity between unlinked nodes. In particular, if a pair of unlinked drug and target in $\bm{A}^{i,j}$ has a certain level of proximity (none zero value) 
in $\bm{M}^{i,j}$, the corresponding relation represented by the DeepWalk matrix could be interpreted as the recovery of their missing interaction, which supplements the incomplete interaction information and reduces the noise in the original dataset. See an example in Supplementary Section A2.1.

In order to mine multiple DeepWalk matrices effectively, we define a unified DeepWalk matrix for the whole DTI network by aggregating every $\bm{M}^{i,j}$:
\begin{equation}
    \bar{\bm{M}} = \sum_{i=1}^{m_d}\sum_{j=1}^{m_t} w^d_i w^t_j \bm{M}^{i,j},
    \label{eq:M_}
\end{equation}
where $w^d_i$ and $w^t_j$ are weights of $i$-th drug and $j$-th target layers respectively with $\sum_{i=1}^{m_d}w^d_i=1$ and $\sum_{j=1}^{m_t}w^t_j=1$. 
In Eq.~\eqref{eq:M_}, the importance of each hyper-layer is determined by multiplying the weights of its involved drug and target layers. This work employs the local interaction consistency (LIC)-based similarity weight, which assesses the proportion of proximate drugs (targets) having the same interactions, and has been found more effective than other similarity weights for DTI prediction~\cite{liu2021optimizing}. More details on LIC weights can be found in Supplementary Section A2.2.  

Let $\bm{Q} = \begin{bmatrix} \bm{U} \\ \bm{V} \end{bmatrix}$ be the vertical concatenation  of drug and target embeddings. We encourage $\bm{Q}\bm{Q}^\top$ to approximate $\bar{\bm{M}}$, which enables the learned embeddings to capture the topology information characterized by the holistic DeepWalk matrix. Hence, we derive the DeepWalk regularization term that diminishes the discrepancy between $\bar{\bm{M}}$ and $\bm{Q}\bm{Q}^\top$:
\begin{equation}
    \mathcal{R}_{dw}(\bm{U},\bm{V}) = ||\bar{\bm{M}}-\bm{Q}\bm{Q}^\top||_F^2.
    \label{eq:M_regularization'}
\end{equation}
Considering that the adjacency matrix $\bm{A}^{i,j}$ includes four blocks, $\bar{\bm{M}}$ and $\bm{Q}\bm{Q}^\top$ could be divided into four blocks accordingly:
\begin{equation}
\bar{\bm{M}} = \begin{bmatrix} 
\bar{\bm{M}}_{S_d} & \bar{\bm{M}}_{Y} \\
{\bar{\bm{M}}_{Y}}^\top & \bar{\bm{M}}_{S_t}  \\
\end{bmatrix} \quad
\bm{Q}\bm{Q}^\top=\begin{bmatrix} 
\bm{U}\bm{U}^\top & \bm{U}\bm{V}^\top  \\
\bm{V}\bm{U}^\top & \bm{V}\bm{V}^\top  \\
\end{bmatrix}.
\end{equation}
Thus, $\mathcal{R}_{dw}(\bm{U},\bm{V})$ can be expressed as the sum of norms of these blocks: 
\begin{equation}
\begin{aligned}
    \mathcal{R}_{dw}(\bm{U},\bm{V}) = &  ||\bar{\bm{M}}_{S_d}-\bm{U}{\bm{U}}^\top||_F^2 + 2||\bar{\bm{M}}_{Y}-\bm{U}{\bm{V}}^\top||_F^2 + \\ 
    & ||\bar{\bm{M}}_{S_t}-\bm{V}{\bm{V}}^\top||_F^2.
 \label{eq:M_regularization}
\end{aligned}
\end{equation}

However, aggregating all per-layer DeepWalk matrices to the holistic one inevitably leads to substantial loss of layer-specific topology information. To address this limitation, we employ graph regularization for each sparsified drug (target) layer to preserve per layer drug (target) proximity in the embedding space, i.e., similar drugs (targets) in each layer are likely to have similar latent features. To distinguish the utility of each layer, each graph regularization is multiplied by the LIC-based weight of its corresponding layer, which emphasizes the proximity of more reliable similarities.
Furthermore, Tikhonov regularization is added to prevent latent features from overfitting the training set. 

By replacing $\mathcal{R}(\bm{U},\bm{V})$ in Eq.~\eqref{eq:MF_obj}  with the above regularization terms, we arrive to the objective of MDMF:
\begin{equation}
\begin{aligned}
    \min_{\bm{U},\bm{V}} \quad & \mathcal{L}(\hat{\bm{Y}},\bm{Y}) + \frac{\lambda_M}{2}\mathcal{R}_{dw}(\bm{U},\bm{V}) + \frac{\lambda_d}{2}\sum_{i=1}^{m_d}w^d_i\text{tr}({\bm{U}}^{\top} \bm{L}^d_i \bm{U})\\ &  +  \frac{\lambda_t}{2}\sum_{j=1}^{m_t}w^t_j\text{tr}({\bm{V}}^\top \bm{L}^t_j \bm{V}) + \frac{\lambda_r}{2}\left(||\bm{U}||_{F}^2+||\bm{V}||_{F}^2 \right), 
\end{aligned}
\label{eq:MDMF_obj}  
\end{equation}
where $\bm{L}^d_i=\text{diag}(\hat{\bm{S}}^{d,i}\bm{e})-\hat{\bm{S}}^{d,i}$ and $\bm{L}^t_j=\text{diag}(\hat{\bm{S}}^{t,j}\bm{e})-\hat{\bm{S}}^{t,j}$ are graph Laplacian matrices of $\hat{\bm{S}}^{d,i}$ and $\hat{\bm{S}}^{t,j}$ respectively, $\lambda_M$, $\lambda_d$, $\lambda_t$, and $\lambda_r$ are regularization coefficients. 
Eq. \eqref{eq:MDMF_obj} can be solved by updating $\bm{U}$ and $\bm{V}$ alternatively ~\cite{Liu2016NeighborhoodPrediction,Ezzat2017Drug-targetFactorization}, using an optimization algorithm, e.g., gradient descent (GD) or AdaGrad~\cite{Duchi2011AdaptiveOptimization}. The details for the optimization procedure of MDMF are provided in Supplementary Section A2.3.

\subsection{Optimizing the Area Under the Curve with MDMF}

\subsubsection{Area Under the Curve Loss Functions}
AUPR and AUC are two widely used area under the curve metrics in DTI prediction. Modeling differentiable surrogate losses that optimize these two metrics can lead to improvements in predicting performance~\cite{liu2021optimizing}. Therefore, we instantiate the loss function in Eq. \eqref{eq:MDMF_obj} with AUPR and AUC losses, and derive two DeepWalk-based MF models, namely MDMFAUPR and MDMFAUC, that optimize the AUPR and AUC metrics, respectively.

Given $\bm{Y}$ and its predictions $\bm{\hat{Y}}=\sigma(\bm{U}\bm{V}^\top)$, which are sorted in descending order according to their predicted scores, AUPR without the interpolation  to estimate the curve is computed as:
\begin{equation}
\begin{aligned}
\text{AUPR}(\bm{\hat{Y}}, \bm{Y}) &= \sum_{h=1}^{n_dn_t} \text{Prec}@h * \text{InRe}@h, \\ 
\end{aligned}
\label{eq:AUPR}
\end{equation}
where \text{Prec}$@h$ is the precision of the $h$ first  predictions, \text{InRe}$@h$ is the incremental recall from rank $h$-1 to $h$. In addition, histogram binning~\cite{Revaud2019LearningLoss}, which assigns predictions into $n_b$ ordered bins, is employed to simulate the non-differentiable and non-smooth sorting operation used to rank predictions, deriving differential precision and incremental recall as:
\begin{equation}
\text{Prec'}@h = \frac{\sum_{i=1}^h \psi(\delta(\hat{\bm{Y}},i)\odot \bm{Y})}{\sum_{i=1}^h \psi(\delta(\hat{\bm{Y}},i))}
\label{eq:Prec}
\end{equation}
\begin{equation}
\text{InRe'}@h = \frac{1}{\psi({\bm{Y}})}\psi(\delta(\hat{\bm{Y}},h)\odot \bm{Y}) ,
\label{eq:InRe}
\end{equation}
where $\delta(\hat{\bm{Y}},h)$ is the soft assignment function which returns the membership degree of each prediction to the $h$-th bin, i.e., $[\delta(\hat{\bm{Y}},h)]_{ij} = \max\left( 1-|\hat{Y}_{ij}-b_h|/\Delta,0 \right)$, where $b_h$ is the center of the $h$-th bin, and $\Delta= 1/(n_b-1)$ is the bin width.
Considering that maximizing AUPR is equivalent to minimizing $-\text{AUPR}(\bm{\hat{Y}}, \bm{Y})$, we obtain the differential AUPR loss according to Eq. \eqref{eq:AUPR}-\eqref{eq:InRe}:
\begin{equation}
\mathcal{L}_{AP} = -\sum_{h=1}^{n_b}\frac{\psi(\delta(\hat{\bm{Y}},h)\odot \bm{Y}) \sum_{i=1}^h \psi(\delta(\hat{\bm{Y}},i)\odot \bm{Y})}{\sum_{i=1}^h  \psi(\delta(\hat{\bm{Y}},i))}.
\label{eq:L_AP}
\end{equation}
The objective of MDMFAUPR is given by using $\mathcal{L}_{AP}$ to replace $\mathcal{L}(\hat{\bm{Y}},\bm{Y})$ in Eq. \eqref{eq:MDMF_obj}.

AUC, which assesses the proportion of correctly ordered tuples, where the interacting drug-target pairs have higher prediction than the non-interacting ones, is defined as:
\begin{equation}
\text{AUC}(\hat{\bm{Y}},\bm{Y}) = \frac{1}{|P|}\sum_{(ij,hl)\in P}\llbracket \hat{Y}_{ij} >\hat{Y}_{hl} \rrbracket,
\label{eq:AUROC}
\end{equation} 
where  predictions $\hat{\bm{Y}}=\bm{U}\bm{V}^\top$ with $f=\omega$, and $P=\{(ij,hl)|(i,j) \in P_1, (h,l) \in P_0\}$ is the Cartesian product of $P_1=\{(i,j)|Y_{ij}=1\}$ and $P_0=\{(i,j)|Y_{ij}=0\}$.
To maximize AUC, we need to minimize the proportion of wrongly ordered tuples, where the prediction of the interacting pair is lower than the non-interacting one. In order to make the AUC loss to be optimized easily, the discontinuous indicator function is substituted by a convex approximate function, i.e., $\phi(x)=\log(1+\exp(-x))$, and the derived convex AUC loss is: 
\begin{equation}
\mathcal{L}_{AUC} = \sum_{(ij,hl)\in P} \phi( \zeta_{ijhl}),
\end{equation}
where $\zeta_{ijhl}=\hat{Y}_{ij}-\hat{Y}_{hl}$.
With the substitution of $\mathcal{L}(\hat{\bm{Y}},\bm{Y})$ with $\mathcal{L}_{AUC}$ in Eq. \eqref{eq:MDMF_obj}, we obtain the objective of MDMFAUC.

More details for optimizing the AUPR and AUC losses can be found in Supplementary Section A2.4.

\subsubsection{Inferring Embeddings of New Entities}

When a new drug (target) arrives, MDMFAUPR and MDMFAUC infer its embeddings using its neighbors in the training set. Given a new drug, we first compute its similarities with all training drugs in $D$ based on its chemical structure, side effects, etc. Then, we linearly integrate the different types of similarity with LIC-based weights to obtain the fused similarity vector:  
\begin{equation}
    \bar{\bm{s}}^{\alpha}_x = \sum_{h=1}^{m_\alpha}w^{\alpha}_h \bar{\bm{s}}^{\alpha,h}_{x}, \quad \alpha \in \{d,t\},
\end{equation}
where $\alpha_x$ is an unseen entity with $\alpha=d$ denoting a new drug and $\alpha=t$ indicating a novel target, and $\bar{\bm{s}}^{\alpha}_x \in \mathbb{R}^{n_\alpha}$ is the fused similarity vector of $\alpha_x$.
Based on $\bar{\bm{s}}^{\alpha}_x$, we retrieve the $k$-NNs of ${\alpha}_x$ from the training set, denoted as $\mathcal{N}^k_{\alpha_x}$, and estimate the embeddings of $\alpha_x$ as follows:
\begin{align}
\bm{U}_x & = \frac{1}{\sum_{d_i \in {\mathcal{N}^k_{d_x}}}\bar{s}^d_{xi}} \sum_{d_i \in {\mathcal{N}^k_{d_x}}}\eta^{i'-1}\bar{s}^d_{xi}\bm{U}_{i}, \quad \text{if } \alpha=d \nonumber \\ 
\bm{V}_x & = \frac{1}{\sum_{t_j \in {\mathcal{N}^k_{t_x}}}\bar{s}^t_{xj}} \sum_{t_j \in {\mathcal{N}^k_{t_x}}}\eta^{j'-1}\bar{s}^t_{xj}\bm{V}_{j}, \quad \text{if } \alpha=t,
\label{eq:combine_test_sim}
\end{align}
where $\bar{s}^d_{xi}$ is the similarity between $d_x$ and $d_i$, $i'$ ($j'$) is the rank of $d_i$ ($t_j$) among $\mathcal{N}^k_{d_x}$ ($\mathcal{N}^k_{t_x}$), e.g., $i'=2$ if $d_i$ is the second nearest training drug of $d_x$, and $\eta \in (0,1]$ is the decay coefficient shrinking the weight of further neighbors, which is an important parameter to control the embedding aggregation.
In addition, we employ a pseudo embedding based $\eta$ selection strategy~\cite{liu2021optimizing} to choose the optimal $\eta$ values from a set of candidate $\mathcal{C}$ for prediction settings involving new drugs or/and new targets, i.e., S2, S3, and S4, respectively. In MDMFAUPR, the $\eta$ value leading to the best AUPR result is used, while the optimal $\eta$ in MDMFAUC is the one achieving the best AUC result.

\subsubsection{MDMF2A: Combining AUPR and AUC}

\begin{figure*}[t]
\centering
\includegraphics[width=0.9\textwidth]{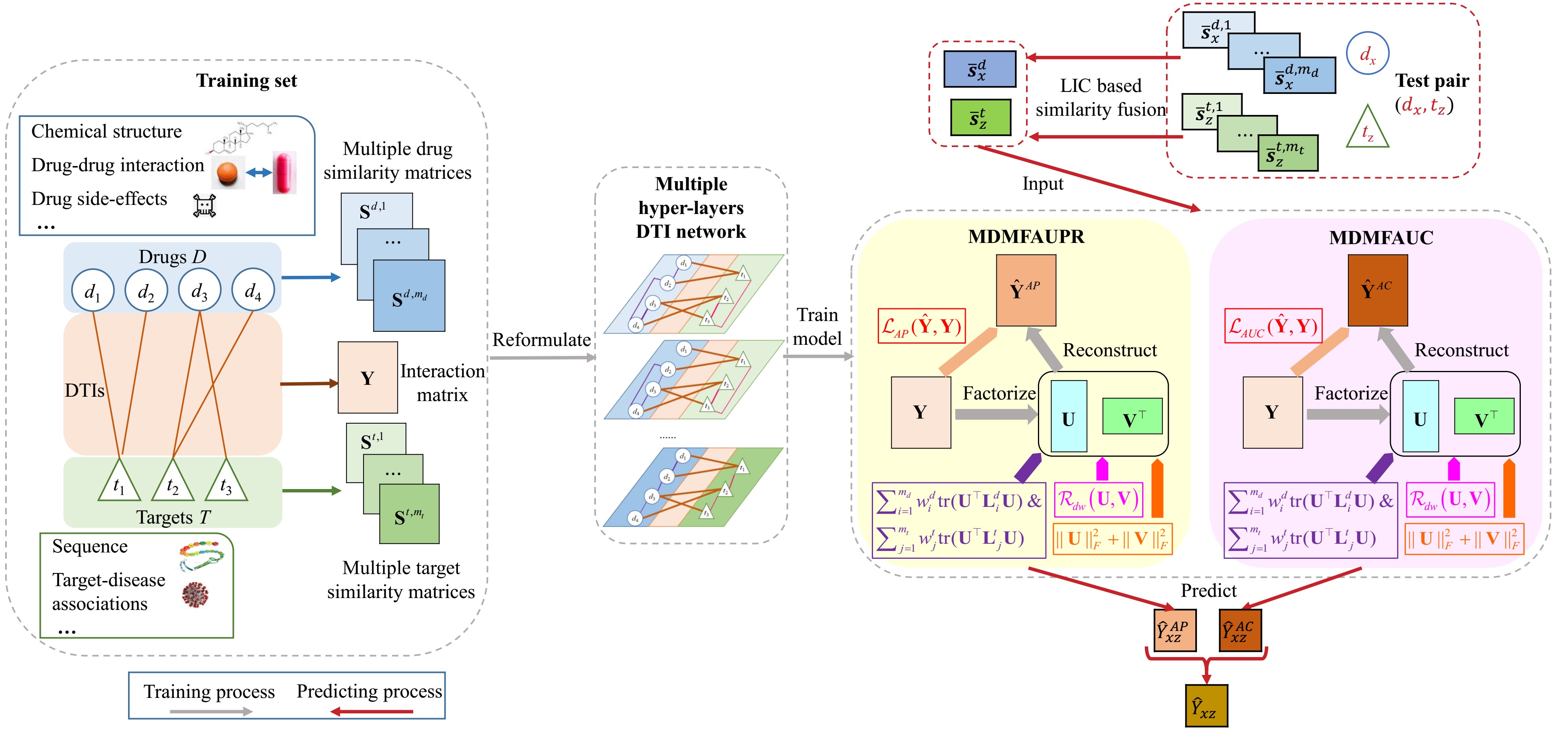}
\caption{The flowchart of MDMF2A. The DTI dataset, consisting of multiple drug and target similarities, is firstly reformulated as a multiplex DTI network containing multiple hyper-layers. Then, two base models of MDMF2A, namely MDMFAUPR and MDMFAUC, are trained upon the derived multiple hyper-layers DTI network with the instantiation of two area under the curves metric based losses, respectively. Then, given a test pair, 
its associated similarity vectors are fused according to the LIC-based weights.
Next, the two base models leverage these fused similarities to generate the estimation of the test pair, respectively. Lastly, MDMF2A aggregates the outputs of two models using Eq. \eqref{eq:MDMF2A_aggregate_prediction} to obtain its final prediction.
}
\label{fig:MDMF2A}
\end{figure*}

It is known that both AUPR and AUC play a vital role in DTI prediction. However, MDMFAUPR and MDMFAUC only optimize one measure but ignore the other. 
To overcome this limitation, we propose an ensemble approach, called DWFM2A, which integrates the two MF models by aggregating their predicted scores. Given a test pair $(d_x, t_z)$, along with its predicted scores $\hat{Y}^{AP}_{xz}$ and $\hat{Y}^{AC}_{xz}$ obtained from MDMFAUPR and MDMFAUC respectively, the final prediction output by MDMF2A is defined as:
\begin{equation}
\hat{Y}_{xz} = \beta \hat{Y}^{AP}_{xz}+(1-\beta)\sigma(\hat{Y}^{AC}_{xz}),
\label{eq:MDMF2A_aggregate_prediction}
\end{equation}
where $\beta \in [0,1]$ is the trade-off coefficient for the two MF base models, and $\sigma$ converts $\hat{Y}^{AC}_{xz}$ and $\hat{Y}^{AP}_{xz}$ into the same scale, i.e., $(0,1)$.
The flowchart of MDMF2A is shown in Figure~\ref{fig:MDMF2A}.

The computational complexity analysis of the proposed methods can be found in Supplementary Section A2.5.

\section{Experimental Evaluation and Discussion}
\label{sec.experiments}



\subsection{Experimental Setup}

\begin{table*}[!t]
\centering
\setlength{\tabcolsep}{2pt}
\caption{AUPR results in all prediction settings.}
\label{tab:result_aupr}
\begin{tabular}{@{}ccccccccccccc@{}}
\toprule
Setting & Dataset & WkNNIR & MSCMF & NRMFL & GRGMF & MF2A & DRLSM & DTINet & NEDTP & NetMF & Multi2Vec & MDMF2A \\ \midrule
\multirow{5}{*}{S1} & NR & - & 0.628(6) & 0.64(5) & 0.658(3) & 0.673(2) & 0.642(4) & 0.508(8) & 0.546(7) & 0.455(9) & 0.43(10) & \textbf{0.675(1)} \\
 & GPCR & - & 0.844(4.5) & 0.86(3) & 0.844(4.5) & 0.87(2) & 0.835(6) & 0.597(10) & 0.798(8) & 0.831(7) & 0.747(9) & \textbf{0.874(1)} \\
 & IC & - & 0.936(3) & 0.934(4) & 0.913(7) & 0.943(2) & 0.914(6) & 0.693(10) & 0.906(8) & 0.929(5) & 0.867(9) & \textbf{0.946(1)} \\
 & E & - & 0.818(6) & 0.843(3) & 0.832(4) & 0.858(2) & 0.811(7) & 0.305(10) & 0.78(8) & 0.831(5) & 0.736(9) & \textbf{0.859(1)} \\
 & Luo & - & 0.599(6) & 0.603(5) & 0.636(3) & 0.653(2) & 0.598(7) & 0.216(9) & 0.08(10) & 0.615(4) & 0.451(8) & \textbf{0.679(1)} \\
\multicolumn{2}{c}{$AveRank$} & - & 5.1 & 4 & 4.3 & 2 & 6 & 9.4 & 8.2 & 6 & 9 & \textbf{1} \\ \midrule
\multirow{5}{*}{S2} & NR & 0.562(5) & 0.531(7) & 0.547(6) & 0.564(4) & 0.578(2) & 0.57(3) & 0.339(10) & 0.486(8) & 0.34(9) & 0.338(11) & \textbf{0.602(1)} \\
 & GPCR & 0.54(4) & 0.472(7) & 0.508(6) & 0.542(3) & 0.551(2) & 0.532(5) & 0.449(9) & 0.451(8) & 0.356(10) & 0.254(11) & \textbf{0.561(1)} \\
 & IC & 0.491(4) & 0.379(8.5) & 0.479(5) & 0.493(3) & 0.495(2) & 0.466(6) & 0.365(10) & 0.407(7) & 0.379(8.5) & 0.16(11) & \textbf{0.502(1)} \\
 & E & 0.405(4) & 0.288(8) & 0.389(5) & 0.415(3) & 0.422(2) & 0.376(6) & 0.173(10) & 0.33(7) & 0.269(9) & 0.141(11) & \textbf{0.428(1)} \\
 & Luo & 0.485(2) & 0.371(8) & 0.458(6) & 0.462(5) & 0.472(4) & \textbf{0.502(1)} & 0.187(9) & 0.077(11) & 0.376(7) & 0.079(10) & 0.48(3) \\
\multicolumn{2}{c}{$AveRank$} & 3.8 & 7.7 & 5.6 & 3.6 & 2.4 & 4.2 & 9.6 & 8.2 & 8.7 & 10.8 & \textbf{1.4} \\ \midrule
\multirow{5}{*}{S3} & NR & 0.56(3) & 0.505(7) & 0.519(6) & 0.545(5) & \textbf{0.588(1)} & 0.546(4) & 0.431(8) & 0.375(9) & 0.359(10) & 0.335(11) & 0.582(2) \\
 & GPCR & 0.774(3) & 0.69(7) & 0.729(6) & 0.755(5) & 0.787(2) & 0.757(4) & 0.546(10) & 0.684(8) & 0.638(9) & 0.327(11) & \textbf{0.788(1)} \\
 & IC & 0.861(3) & 0.827(7) & 0.838(6) & 0.851(5) & 0.863(2) & 0.855(4) & 0.599(10) & 0.8(8) & 0.791(9) & 0.4(11) & \textbf{0.865(1)} \\
 & E & 0.728(3) & 0.623(8) & 0.711(6) & 0.715(4) & 0.731(2) & 0.712(5) & 0.313(10) & 0.615(9) & 0.656(7) & 0.257(11) & \textbf{0.738(1)} \\
 & Luo & 0.243(5) & 0.08(8) & 0.204(7) & 0.234(6) & 0.292(3) & 0.248(4) & 0.061(9) & 0.046(10) & 0.294(2) & 0.027(11) & \textbf{0.299(1)} \\
\multicolumn{2}{c}{$AveRank$} & 3.4 & 7.4 & 6.2 & 5 & 2 & 4.2 & 9.4 & 8.8 & 7.4 & 11 & \textbf{1.2} \\ \midrule
\multirow{5}{*}{S4} & NR & \textbf{0.309(1)} & 0.273(6) & 0.278(5) & 0.308(2) & 0.289(3) & 0.236(8) & 0.249(7) & 0.16(9) & 0.146(11) & 0.149(10) & 0.286(4) \\
 & GPCR & 0.393(3) & 0.323(6) & 0.331(5) & 0.368(4) & \textbf{0.407(1.5)} & 0.077(10) & 0.306(7) & 0.29(8) & 0.159(9) & 0.074(11) & \textbf{0.407(1.5)} \\
 & IC & 0.339(4) & 0.194(9) & 0.327(5) & 0.347(3) & 0.352(2) & 0.086(10) & 0.264(6) & 0.251(7) & 0.196(8) & 0.067(11) & \textbf{0.356(1)} \\
 & E & 0.228(3) & 0.074(9) & 0.221(5) & 0.224(4) & 0.235(2) & 0.063(10) & 0.1(7) & 0.112(6) & 0.09(8) & 0.017(11) & \textbf{0.239(1)} \\
 & Luo & 0.132(3) & 0.018(10) & 0.096(6) & 0.106(5) & 0.175(2) & 0.061(7) & 0.035(8) & 0.03(9) & 0.127(4) & 0.002(11) & \textbf{0.182(1)} \\
\multicolumn{2}{c}{$AveRank$} & 2.8 & 8 & 5.2 & 3.6 & 2.1 & 9 & 7 & 7.8 & 8 & 10.8 & \textbf{1.7} \\ \midrule
\multicolumn{2}{c}{$Summary$} & 3.33* & 7.05* & 5.25* & 4.13* & 2.13* & 5.85* & 8.85* & 8.25* & 7.53* & 10.4* & \textbf{1.33} \\ \bottomrule
\end{tabular}
\end{table*}

Following~\cite{Pahikkala2015TowardPredictions}, four types of cross validation (CV) are conducted to examine the methods in four prediction settings, respectively. In S1, the 10-fold CV on pairs is used, where one fold of pairs is removed for testing. In S2 (S3), the 10-fold CV on drugs (targets) is applied, where one drug (target) fold along with its corresponding rows (columns) in $\bm{Y}$ is separated for testing. The 3$\times$3-fold block-wise CV, which splits a drug fold and target fold along with the interactions between them for testing, using the interactions between the remaining drugs and targets for training, is applied to S4. AUPR and AUC defined in Eq. \eqref{eq:AUPR} and \eqref{eq:AUROC} are used as evaluation measures.

To evaluate the performance of MDMF2A in all prediction settings, we have compared it to seven DTI prediction models (WkNNIR~\cite{Liu2021Drug-targetRecovery}, NRLMF~\cite{Liu2016NeighborhoodPrediction}, MSCMF~\cite{Zheng2013CollaborativeInteractions}, GRGMF~\cite{Zhang2020ANetworks}, MF2A~\cite{liu2021optimizing}, DRLSM~\cite{Ding2020IdentificationFusion}, DTINet~\cite{Luo2017AInformation}, NEDTP~\cite{An2021AInteractions})
and two network embedding approaches applicable to any domain (NetMF~\cite{qiu2018network} and Multi2Vec~\cite{Teng2021AEmbedding}).
WkNNIR cannot perform predictions in S1, as it is specifically designed to predict interactions involving new drugs or/and targets (S2, S3, S4)~\cite{Liu2021Drug-targetRecovery}.
Furthermore, the proposed MDMF2A is also compared to four deep learning-based methods, namely NeoDTI~\cite{Wan2019NeoDTI:Interactions}, DTIP~\cite{Xuan2021IntegratingPrediction}, DCFME~\cite{chen2022DCFME}, and SupDTI~\cite{chen2022SupDTI}. These deep learning competitors can only be applied to the \textsl{Luo} dataset in S1, because they formulate the DTI dataset as a heterogeneous network consisting of four types of nodes (drugs, targets, drug-side effects, and diseases) and learn embeddings for all types of nodes.
The illustration of all baseline methods can be found in Supplementary Section A3.

The parameters of all baseline methods are set based on the suggestions in the respective articles. For MDMF2A, the trade-off ensemble weight $\beta$ is chosen from $\{0,0.01,\ldots,1.0\}$. For the two base models of MDMF2A, i.e., MDMFAUPR and MDMFAUC, the number of neighbors is set to $k=5$, the window size of the random walk to $n_w=5$, the number of negative samples to $n_s=1$, the learning rate to $\theta=0.1$, candidate decay coefficient set $\mathcal{C}$=\{0.1, 0.2, $\dotsc$, 1.0\}, the embedding dimension $r$ is chosen from \{50, 100\}, $\lambda_d$, $\lambda_t$ and $\lambda_r$ are chosen from \{$2^{-6}$, $2^{-4}$, $2^{-2}$, $2^{0}$, $2^{2}$\}, and $\lambda_M=0.005$. The number of bins $n_b$ in MDMFAUPR is chosen from \{11, 16, 21, 26, 31\} for small and medium datasets (\textsl{NR}, \textsl{GPCR}, and \textsl{IC}), while it is set to 21 for larger datasets (\textsl{E} and \textsl{Luo}). Similar to ~\cite{Zheng2013CollaborativeInteractions,Liu2016NeighborhoodPrediction,Ding2020IdentificationFusion}, we obtain the best hyperparameters of our model via grid search, and the detailed settings are listed in Supplementary Table A2. 

\subsection{Results and Discussion}

Tables \ref{tab:result_aupr} and \ref{tab:result_auc} list the results of MDMF2A and its nine competitors on five datasets under four prediction settings. The numbers in the last row denote the average rank across all prediction settings, and ``*" indicates that the advantage of DAMF2A over the competitor is statistically significant according to a Wilcoxon signed-rank test with Bergman-Hommel’s correction~\cite{Benavoli2016} at 5\% level on the results of all prediction settings.

\begin{table*}[t]
\setlength{\tabcolsep}{2pt}
\caption{AUC results in all prediction settings.}
\label{tab:result_auc}
\begin{tabular}{@{}ccccccccccccc@{}}
\toprule
Setting & Dataset & WkNNIR & MSCMF & NRMFL & GRGMF & MF2A & DRLSM & DTINet & NEDTP & NetMF & Multi2Vec & MDMF2A \\ \midrule
\multirow{5}{*}{S1} & NR & - & 0.882(4.5) & 0.882(4.5) & 0.891(2) & 0.884(3) & 0.879(6) & 0.797(9) & 0.846(7) & 0.818(8) & 0.788(10) & \textbf{0.892(1)} \\
 & GPCR & - & 0.962(6) & 0.972(4) & \textbf{0.978(2)} & \textbf{0.978(2)} & 0.971(5) & 0.916(10) & 0.953(8) & 0.96(7) & 0.93(9) & \textbf{0.978(2)} \\
 & IC & - & 0.982(6) & \textbf{0.989(2)} & 0.988(4) & \textbf{0.989(2)} & 0.981(7.5) & 0.938(10) & 0.981(7.5) & 0.985(5) & 0.97(9) & \textbf{0.989(2)} \\
 & E & - & 0.961(8) & 0.981(4) & 0.982(3) & 0.983(2) & 0.964(7) & 0.839(10) & 0.97(5) & 0.966(6) & 0.944(9) & \textbf{0.984(1)} \\
 & Luo & - & 0.922(7) & 0.951(3) & 0.947(4) & 0.966(2) & 0.941(5) & 0.894(9) & 0.929(6) & 0.917(8) & 0.861(10) & \textbf{0.97(1)} \\
\multicolumn{2}{c}{$AveRank$} & - & 6.3 & 3.5 & 3 & 2.2 & 6.1 & 9.6 & 6.7 & 6.8 & 9.4 & \textbf{1.4} \\ \hline
\multirow{5}{*}{S2} & NR & 0.825(5.5) & 0.802(7) & 0.826(4) & 0.825(5.5) & 0.833(2) & 0.831(3) & 0.666(11) & 0.786(8) & 0.739(9) & 0.727(10) & \textbf{0.837(1)} \\
 & GPCR & 0.914(4) & 0.882(7) & 0.913(5) & 0.924(2.5) & 0.924(2.5) & 0.867(8) & 0.858(9) & 0.885(6) & 0.852(10) & 0.811(11) & \textbf{0.925(1)} \\
 & IC & 0.826(4) & 0.783(9) & 0.825(5) & \textbf{0.833(1)} & 0.828(2) & 0.796(7) & 0.766(10) & 0.794(8) & 0.803(6) & 0.715(11) & 0.827(3) \\
 & E & 0.877(4) & 0.835(7) & 0.858(5) & 0.891(3) & 0.892(2) & 0.799(9) & 0.78(10) & 0.837(6) & 0.811(8) & 0.732(11) & \textbf{0.895(1)} \\
 & Luo & 0.904(5) & 0.897(7) & 0.917(3) & 0.899(6) & 0.927(2) & 0.864(9) & 0.873(8) & 0.907(4) & 0.861(10) & 0.776(11) & \textbf{0.937(1)} \\
\multicolumn{2}{c}{$AveRank$} & 4.5 & 7.4 & 4.4 & 3.6 & 2.1 & 7.2 & 9.6 & 6.4 & 8.6 & 10.8 & \textbf{1.4} \\ \hline
\multirow{5}{*}{S3} & NR & 0.82(3) & 0.786(7) & 0.825(2) & \textbf{0.845(1)} & 0.819(4) & 0.798(6) & 0.756(8) & 0.726(9) & 0.712(10) & 0.703(11) & 0.805(5) \\
 & GPCR & 0.952(4) & 0.902(9) & 0.946(5) & \textbf{0.965(1)} & 0.96(3) & 0.917(7) & 0.879(10) & 0.918(6) & 0.909(8) & 0.798(11) & 0.961(2) \\
 & IC & 0.958(5) & 0.941(7) & 0.96(4) & 0.965(3) & \textbf{0.967(1)} & 0.942(6) & 0.907(10) & 0.939(8) & 0.938(9) & 0.866(11) & 0.966(2) \\
 & E & 0.935(5) & 0.881(9) & 0.936(4) & 0.943(3) & 0.944(2) & 0.883(8) & 0.841(10) & 0.918(6) & 0.911(7) & 0.815(11) & \textbf{0.948(1)} \\
 & Luo & 0.835(5) & 0.826(9) & 0.828(8) & 0.84(4) & 0.901(2) & 0.801(10) & 0.829(7) & 0.86(3) & 0.831(6) & 0.633(11) & \textbf{0.902(1)} \\
\multicolumn{2}{c}{$AveRank$} & 4.4 & 8.2 & 4.6 & 2.4 & 2.4 & 7.4 & 9 & 6.4 & 8 & 11 & \textbf{2.2} \\ \hline
\multirow{5}{*}{S4} & NR & 0.637(5) & 0.597(6) & 0.656(3) & \textbf{0.677(1)} & 0.649(4) & 0.592(7) & 0.562(8) & 0.548(9) & 0.531(10) & 0.524(11) & 0.661(2) \\
 & GPCR & 0.871(4) & 0.798(8) & 0.866(5) & \textbf{0.89(1)} &0.886(3) & 0.405(11) & 0.803(7) & 0.816(6) & 0.721(9) & 0.581(10) & 0.887(2) \\
 & IC & 0.774(5) & 0.658(9) & 0.775(4) & 0.782(2) & 0.776(3) & 0.498(11) & 0.706(6) & 0.702(7) & 0.683(8) & 0.555(10) & \textbf{0.783(1)} \\
 & E & 0.819(3) & 0.695(8) & 0.799(5) & 0.815(4) & 0.821(2) & 0.453(11) & 0.757(6) & 0.744(7) & 0.69(9) & 0.541(10) & \textbf{0.827(1)} \\
 & Luo & 0.819(4) & 0.752(7) & 0.804(5) & 0.745(8) & 0.848(2) & 0.438(11) & 0.787(6) & 0.822(3) & 0.732(9) & 0.513(10) & \textbf{0.85(1)} \\
\multicolumn{2}{c}{$AveRank$} & 4.2 & 7.6 & 4.4 & 3.2 & 2.8 & 10.2 & 6.6 & 6.4 & 9 & 10.2 & \textbf{1.4} \\ \hline
\multicolumn{2}{c}{$Summary$} & 4.37* & 7.38* & 4.23* & 3.05 & 2.38* & 7.73* & 8.7* & 6.48* & 8.1* & 10.35* & \textbf{1.6} \\ \bottomrule
\end{tabular}
\end{table*}

The proposed MDMF2A is the best performing model in most cases for both metrics, achieving the highest average rank in all prediction settings and statistically significantly outperforming all competitors, except for GRGMF in AUC. This demonstrates the effectiveness of our model to sufficiently exploit the topology information embedded in the multiplex heterogeneous DTI network and optimize the two area under the curve metrics. 
MF2A is the runner-up. Its inferiority to MDMF2A is mainly attributed to the ignorance of high-order proximity captured by random walks and the view-specific information loss caused by aggregating multi-type similarities.
GRGMF, WkNNIR, NRMLF, DRLSM, and MSCMF come next. They are outperformed by the proposed MDMF2A, because they fail to capture the unique information provided by each view. 
The two graph embedding based DTI prediction models are usually inferior to other DTI prediction approaches, because they generate embedding in an unsupervised manner without exploiting the interacting information. Specifically, NEDTP using the class imbalance resilient GBDT as the predicting classifier outperforms DTINet which employs simple linear projection to estimate DTIs.  
Regarding the two general multiplex network embedding methods, NetMF is better than Multi2Vec, because the latter requires dichotomizing the edge weights, which wipes out the different influence of the connected nodes in the similarity subnetwork. In addition,  averaging all per-layer embeddings does not distinguish the importance of each hyper-layer, unlike the holistic DeepWalk Matrix used in NetMF, which contributes to the inferiority of Multi2Vec to NetMF as well.

There are some cases where MDMF2A does not achieve the best performance.
Some baseline models, such as WkNNIR, NRLMF, GRGMF and MF2A, are better than MDMF2A on \textsl{NR} datasets, implying that the random walk based embedding generation may not yield enough benefit in the case of small-sized datasets. Besides, concerning the \textsl{Luo} dataset under S2, MDMF2A is outperformed by WKNNIR and DRLSM, which incorporate the neighborhood interaction recovery procedure, in terms of AUPR. In the \textsl{Luo} dataset, the neighbor drugs are more likely to share the same interactions, leading to the effectiveness of neighborhood-based interaction estimation for new drugs. Nevertheless, the AUC results of the two baselines are 3.7\% and 8.4\% lower than DWFM2A, respectively. Also, MF2A is slightly better than MDMF2A on the \textsl{IC} dataset in terms of AUC under S2 and S3, but the gap of results between them is tiny, e.g., 0.001. 
Finally, GRGMF achieves better AUC results than MDMF2A on the \textsl{GPCR} dataset under S3 and S4 as well as the \textsl{IC} dataset under S2, mainly because GRGMF learns neighbour information adaptively. But it is worse than MDMF2A in terms of AUPR, which is more informative when evaluating a model under extremely imbalanced class distribution (sparse interaction).

The advantage of MDMF2A is also observed in the comparison with deep learning based DTI prediction models on the \textsl{Luo} dataset. As shown in Table \ref{tab:result_luo}, MDMF2A outperforms all competitors in terms of AUPR, achieving 14\% improvements over the best-performing competitor (DCFME). In terms of AUC, MDMF2A is only 1.1\% lower than DTIP, and 5.3\%, 3.6\% and 3.9\% higher than NeoDTI, DCFME and, SupDTI. 
Although DTIP emphasizes the AUC performance and slightly outperforms MDMF2A, it suffers a significant decline in the AUPR results. Compared to deep learning competitors, MDMF2A takes full advantage of the information shared by high-order neighbors and explicitly optimizes AUPR and AUC that are more effective than the conventional entropy and focal losses to identifying the less frequent interacting pairs, resulting in better performance.

\begin{table}[!t]
\centering
\caption{Results of MDMF2A and \textit{Deep Learning} models on Luo dataset in S1.}
\label{tab:result_luo}
\begin{tabular}{@{}cccccc@{}}
\toprule
Metric & NeoDTI & DTIP & DCFME & SupDTI & MDMF2A \\ \midrule
AUPR & 0.573(4) & 0.399(5) & 0.596(2) & 0.585(3) & \textbf{0.679(1)} \\
AUC & 0.921(5) & \textbf{0.981(1)} & 0.936(3) & 0.933(4) & 0.97(2) \\ \bottomrule
\end{tabular}
\end{table}

To comprehensively investigate the proposed MDMF2A, we conduct an ablation study to demonstrate the effectiveness of its ensemble framework and all regularization terms and analyze the sensitivity of three important parameters, i.e., $r$, $n_w$ and $n_s$. Please see more details in Supplementary Sections A5-A6.

\subsection{Discovery of Novel DTIs}

We examine the capability of MDMF2A to find novel DTIs not recorded in the \textsl{Luo} dataset. We do not consider updated golden standard datasets, since they have included all recently validated DTIs collected from up-to-date databases.
We split all non-interacting pairs into 10 folds, and obtain predictions of each fold by training an MDMF2A model with all interacting pairs and the other nine folds of non-interacting ones. All non-interacting pairs are ranked based on their predicting scores, and the top ten pairs are selected as newly discovered DTIs, which are shown in Table \ref{tab:new_DTIs}. 
To further verify the reliability of these new interaction candidates, we search their supportive evidence from DrugBank (DB)~\cite{Wishart2018DrugBank2018} and DrugCentral (DC)~\cite{Avram2021DrugCentralRepositioning}. As we can see, 8/10 new interactions (in bold) are confirmed, demonstrating the success of MDMF2A in trustworthy new DTI discovery.

\begin{table}[h]
\centering
\small
\setlength{\tabcolsep}{2pt}
\caption{Top 10 new DTIs discovered by MDMF2A from \textsl{Luo}’s datasets.}
\label{tab:new_DTIs}
\begin{tabular}{cccccc}
\toprule
Drug ID & Drug name & Target ID & Target name & Rank & Database \\ \midrule
\textbf{DB00829} & \textbf{Diazepam} & \textbf{P48169} & \textbf{GABRA4} & \textbf{1} & \textbf{DB} \\
\textbf{DB01215} & \textbf{Estazolam} & \textbf{P48169} & \textbf{GABRA4} & \textbf{2} & \textbf{DB} \\
DB00580 & Valdecoxib & P23219 & PTGS1 & 3 & - \\
\textbf{DB01367} & \textbf{Rasagiline} & \textbf{P21397} & \textbf{MAOA} & \textbf{4} & \textbf{DC} \\
\textbf{DB00333} & \textbf{Methadone} & \textbf{P41145} & \textbf{OPRK1} & \textbf{5} & \textbf{DC} \\
\textbf{DB00363} & \textbf{Clozapine} & \textbf{P21918} & \textbf{DRD5} & \textbf{6} & \textbf{DC} \\
\textbf{DB06216} & \textbf{Asenapine} & \textbf{P21918} & \textbf{DRD5} & \textbf{7} & \textbf{DB} \\
\textbf{DB06800} & \begin{tabular}[c]{@{}c@{}}\textbf{Methylnal} \\ \textbf{-trexone}\end{tabular} & \textbf{P41143} & \textbf{OPRD1} & \textbf{8} & \textbf{DC} \\
DB00802 & Alfentanil & P41145 & OPRK1 & 9 & - \\
\textbf{DB00482} & \textbf{Celecoxib} & \textbf{P23219} & \textbf{PTGS1} & \textbf{10} & \textbf{DC} \\ \bottomrule
\end{tabular}
\end{table}

\section{Conclusion}
\label{sec.conclusions}
This paper proposed MDMF2A, a random walk and matrix factorization based model, to predict DTIs by effectively mining topology information from the multiplex heterogeneous network involving diverse drug and target similarities. It integrates two base predictors that leverage our designed objective function, encouraging the learned embeddings to preserve holistic network and layer-specific topology structures. The two base models utilize the convex AUPR and AUC losses in their objectives, enabling MDMF2A to simultaneously optimize two crucial metrics in the DTI prediction task. 
We have conducted extensive experiments on five DTI datasets under various prediction settings. The results affirmed the superiority of the proposed MDMF2A to other competing DTI prediction methods. Furthermore, the practical ability of MDMF2A to discover novel DTIs was supported by the evidence from online biological databases.

In the future, we plan to extend our model to handle attributed DTI networks, including both topological and feature information for drugs and targets. 

\hspace*{\fill} 

\noindent\fbox{\parbox{0.46\textwidth}{\textbf{Key Points}:
\begin{itemize}
    \item Incorporating multiple hyper-layers based DeepWalk matrix decomposition and layer-specific graph Laplacian to learn robust node representations that preserve both global and view-specific topology. 
    \item MDMF integrates multiplex heterogeneous network representation learning and DTI prediction into a unified optimization framework, learning latent features in a supervised manner and implicitly recoveries possible missing interactions.
    \item MDMF2A, the instantiation of MDMF, optimizes both AUPR and AUC metrics.
    \item Our method statistically significantly outperforms state-of-the-art methods under various prediction settings and can discover new reliable DTIs.
\end{itemize}}}





\section*{Data and Code availability}
The source code and data are available could be found at \href{https://github.com/intelligence-csd-auth-gr/DTI_MDMF2A}{https://github.com/intelligence-csd-auth-gr/DTI\_MDMF2A}


\section*{Funding}
This work was supported by the China Scholarship Council (CSC) [201708500095]; and the French National Research Agency (ANR) under the JCJC project GraphIA [ANR-20-CE23-0009-01].

\section{Biography}
\textbf{Bin Liu} is a lecturer at Key Laboratory of Data Engineering and Visual Computing, Chongqing University of Posts and Telecommunications and received his PhD Degree in computer science from Aristotle University of Thessaloniki. His research interests include multi-label learning and bioinformatics.

\textbf{Dimitrios Papadopoulos} is a PhD student at the School of Informatics, Aristotle University of Thessaloniki. His research interests include supervised machine learning, graph mining, and drug discovery.

\textbf{Fragkiskos D. Malliaros} is an Assistant Professor at Paris-Saclay University, CentraleSupélec and associate researcher at Inria Saclay. His research interests include graph mining, machine learning, and graph-based information extraction.

\textbf{Grigorios Tsoumakas} is an Associate Professor at the Aristotle University of Thessaloniki. His research interests include machine learning (ensembles, multi-target prediction) and natural language processing (semantic indexing, keyphrase extraction, summarization)
 
\textbf{Apostolos N. Papadopoulos} is Associate Professor at the School of Informatics, Aristotle University of Thessaloniki. His research interests include data management, data mining and big data analytics.

\bibliographystyle{unsrt}
\bibliography{BL_copy,ref}

\end{document}


\maketitle

\renewcommand{\thesection}{A\arabic{section}}
\captionsetup[table]{name={Supplementary Table},labelsep=colon}
\renewcommand{\thetable}{A\arabic{table}}
\captionsetup[figure]{name={Supplementary Figure},labelsep=colon}
\renewcommand{\thefigure}{A\arabic{figure}}

\section{Collecting New DTIs for Golden Standard Datasets}
The original golden standard datasets only record the DTIs discovered before they were constructed (in 2008). There are many new DTIs verified after 2008 but not included by original golden standard datasets, leading to the existence of numerous false negative samples (missing interactions). To complete the interaction information of the original golden standard datasets, we added new discovered DTIs reported in the last version of KEGG~\cite{Kanehisa2017KEGG:Drugs}, DrugBank~\cite{Wishart2018DrugBank2018}, and ChEMBL~\cite{Mendez2019ChEMBL:Data} databases. Specifically, we search every non-interacting drug-target pair in the four original datasets to check if its interaction is recorded in the three databases, respectively. For the ChEMBL database, a drug and a target are considered as an interacting pair if the pChembl Value or the negative logarithm transformed stand\_value ($-\log_{10}(\text{stand\_value})$) of their corresponding assay is larger than 5.0~\cite{mayr2018ChEMBLdataset}. 
After the removal of duplicated DTIs found in more than one databases, 
76, 461, 855, 1330 new interactions are appended to original NR, GPCR, IC and E datasets, respectively. As shown in Table \ref{tab:num_DTIs}, there are 166, 1096, 2331, 4256 interactions in the four updated golden standard datasets, respectively.

\begin{table}[h]
\centering
\caption{The Statistic of DTIs in original and updated golden standard datasets, where All denotes the total number of new DTIs added after deleting repeating interactions.}
\label{tab:num_DTIs}
\begin{tabular}{@{}ccccccc@{}}
\toprule
Dataset & Original & Updated & KEGG & DrugBank & ChEMBL & All \\ \midrule
NR & 90 & 166 & 7 & 37 & 41 & 76 \\
GPCR & 635 & 1096 & 136 & 363 & 26 & 461 \\
IC & 1476 & 2331 & 403 & 590 & 38 & 855 \\
E & 2926 & 4256 & 173 & 850 & 541 & 1330 \\ \bottomrule
\end{tabular}
\end{table}

\section{Supplementary Description of Proposed Method}

\subsection{Example of richer node proximity captured by the DeepWalk matrix}
\begin{figure}[h]
\centering
\includegraphics[width=0.8\textwidth]{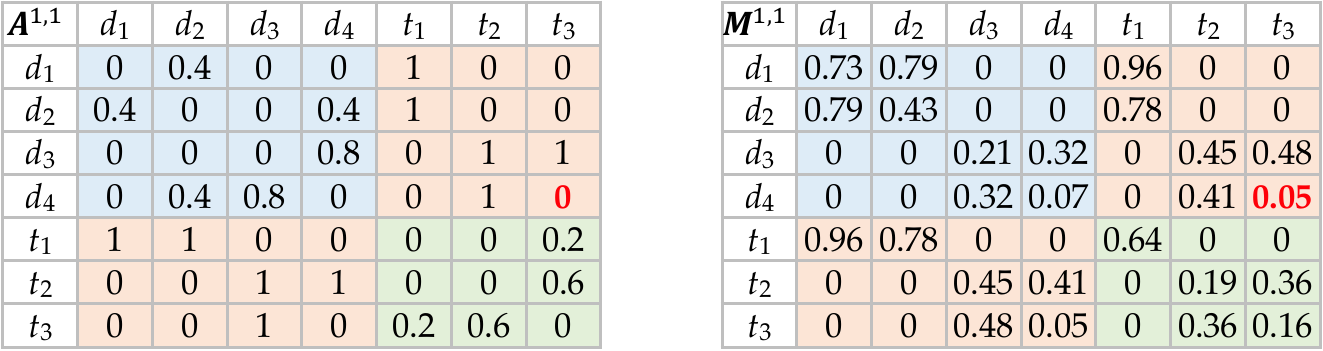}
\caption{Adjacency matrix $\bm{A}^{1,1}$ of Figure 1 and its DeepWalk matrix $\bm{M}^{1,1}$ with $n_w=4$ and $n_s=1$.}  
\label{fig:Compare_A_M}
\end{figure}

In the example shown in Figure~\ref{fig:Compare_A_M}, $d_4$ and $t_3$ are two unlinked nodes in the original network, i.e., there is no bipartite edge connecting them ($A^{1,1}_{4,3}=0$). On the other hand, $M^{1,1}_{4,3}$ is a non-zero value; this indicates that a certain level of proximity between $d_4$ and $t_3$ is found by the DeepWalk matrix, since there are random walk patches with length not exceeding four, such as $d_4 \rightarrow d_3 \rightarrow t_3$, traversing the two nodes. In addition, the relation between $d_4$ and $t_3$ represented by the DeepWalk matrix could also be interpreted as the recovery of their missing interaction, which supplements the incomplete interaction information and reduces the noise in the original dataset. 

\subsection{Local Interaction Consistency Based Similarity Weights}

Most DTI prediction models rely on the \textit{guilt-by-association} assumption~\cite{Olayan2018DDR:Approaches}, i.e., similar drugs tend to interact with the same targets and vice versa. Therefore, the more proximate drugs or targets having identical or similar interactions in a similarity space, the more reliable new DTIs inferred from interaction profiles of neighbourhoods identified by this similarity space. Local Interaction Consistency (LIC) is a measure to assess how is a similarity matrix aligned to \textit{guilt-by-association} rule, which has been shown to have superior performance compared to other similarity weighting strategies by emphasizing the crucial local interaction distribution~\cite{liu2021optimizing}.


Specifically, the LIC of a drug $d_i$ for an interacting target $t_j$ based on the $h$-th drug similarity $S^{d,h}_{i}$ is defined as the proportion of drugs in the local region of $d_i$ having the same interactivity w.r.t. $t_j$: 
\begin{equation}
    {B}^{d,h}_{ij} = 
    \frac{1}{\sum_{d_l\in\mathcal{N}^{k,h}_{d_i}}{S^{d,h}_{il}}}
    \sum_{d_l\in\mathcal{N}^{k,h}_{d_i}}{S^{d,h}_{il}}
    \llbracket {Y}_{il} = {Y}_{ij} \rrbracket,
\label{eq:Bdh_ij}
\end{equation}
where $\mathcal{N}^{k,h}_{d_i}$ denotes the $k$-nearest neighbours of $d_i$ based on $\bm{S}^{d,h}_{i}$, $\llbracket \cdot \rrbracket$ denotes the indicator function returning 1 when the input is true and 0 if it is false, and $Y_{ij}$ indicates the binary interaction value for $d_i$ and $t_j$. To obtain the LIC of a whole similarity matrix $\bm{S}^{d,h}$, we average ${B}^{d,h}_{ij}$ of all drugs with respect to their interacting targets, as shown in Eq. \eqref{eq:LC}, with $P_1 = \{(i,j) | Y_{ij} = 1\}$ denoting all interacting drug-target pairs.
\begin{equation}
    {LC}^{d}_{h} = 
    \frac{1}{|{P}_{1}|}
    \sum_{(i,j)\in{P}_{1}} {B^{d,h}_{ij}}.
\label{eq:LC}
\end{equation}
A higher ${LC}^{d}_{h}$ value represents that more neighbour drugs possessing same or similar interactions in $\bm{S}^{d,h}$, which further indicates the larger reliability of using the neighbourhood interaction information provided by $\bm{S}^{d,h}$ to infer new DTIs.
Finally, we normalize the ${LC}^{d}_{h}$ values for all drug similarities to obtain the LIC-based similarity weights, ensuring that the sum of all similarity weights is 1:
\begin{equation}
    {w}^{d}_{h} = 
    \frac{{LC}^{d}_{h}}{\sum_{i=1}^{m_d} {LC}^{d}_{i}}
    \label{eq:wd_h},
\end{equation}
where $m_d$ is the number of drug similarity matrices.

The LIC based weight of a target similarity $w^t_h$ is calculated in the same way, except for replacing $Y_{il}$ with $Y_{il}^\top$ in Eq. \eqref{eq:Bdh_ij} and substituting 
the superscript $d$ with $t$ in Eq. \eqref{eq:Bdh_ij} - \eqref{eq:wd_h}.

\subsection{MDMF optimization}
The objective of MDMF is defined as:
\begin{equation}
\begin{aligned}
    \min_{\bm{U},\bm{V}} \quad & \mathcal{L}(\hat{\bm{Y}},\bm{Y}) + \frac{\lambda_M}{2}\mathcal{R}_{dw}(\bm{U},\bm{V}) + \frac{\lambda_d}{2}\sum_{i=1}^{m_d}w^d_i\text{tr}({\bm{U}}^{\top} \bm{L}^d_i \bm{U})\\ &  +  \frac{\lambda_t}{2}\sum_{j=1}^{m_t}w^t_j\text{tr}({\bm{V}}^\top \bm{L}^t_j \bm{V}) + \frac{\lambda_r}{2}\left(||\bm{U}||_{F}^2+||\bm{V}||_{F}^2 \right), 
\end{aligned}
\label{eq:MDMF_obj}  
\end{equation}
This optimization problem can be solved by alternating optimization~\cite{Liu2016NeighborhoodPrediction,Ezzat2017Drug-targetFactorization}. In each iteration, $\bm{U}$ is firstly updated with fixed $\bm{V}$, and then $\bm{V}$ is updated with fixed $\bm{U}$.
Let $\mathcal{J}$ denote the objective function to be minimized in Eq. \eqref{eq:MDMF_obj}.  Its gradients w.r.t. $\bm{U}$ and $\bm{V}$ are:
\begin{equation}
\begin{aligned}
    \nabla_{\bm{U}}\mathcal{J} & = \nabla_{\bm{U}}\mathcal{L}(\hat{\bm{Y}},\bm{Y}) +  \lambda_d\sum_{i=1}^{m_d}w^d_i\bm{L}_i^d\bm{U} + \lambda_r\bm{U} + 2\lambda_M\left((\bm{U}{\bm{U}}^\top-\bar{\bm{M}}_{S_d})\bm{U}+(\bm{U}{\bm{V}}^\top-\bar{\bm{M}}_{Y})\bm{V}\right) 
    \label{eq:deriv_U}
\end{aligned}
\end{equation}
\begin{equation}
\begin{aligned}
    \nabla_{\bm{V}}\mathcal{J} & = \nabla_{\bm{V}}\mathcal{L}(\hat{\bm{Y}},\bm{Y}) + \lambda_t\sum_{j=1}^{m_t}w^t_j\bm{L}_j^t\bm{V} + \lambda_r\bm{V} + 2\lambda_M\left((\bm{V}{\bm{V}}^\top-\bar{\bm{M}}_{S_t})\bm{V}+(\bm{V}{\bm{U}}^\top-{\bar{\bm{M}}_{Y}}^\top)\bm{U}\right),
     \label{eq:deriv_V}
\end{aligned}
\end{equation}
where $\nabla_{\bm{U}}\mathcal{L}(\hat{\bm{Y}},\bm{Y})$ and $\nabla_{\bm{V}}\mathcal{L}(\hat{\bm{Y}},\bm{Y})$ are the gradients of the loss function for the two kinds of embeddings, respectively. 
The optimization algorithm of MDMF is illustrated in Algorithm \ref{alg:MDMF}, where $\theta$ is the learning rate.  
We firstly initialize $\bm{U}$ and $\bm{V}$ (Line 1), and calculate some auxiliary variables (Lines 2-6). Then, given an optimization method, which is either gradient descent (GD) or AdaGrad~\cite{Duchi2011AdaptiveOptimization}, we repeat the following procedure: changing $\bm{U}$ based on $\nabla_{\bm{U}}\mathcal{J} $ and subsequently updating $\bm{V}$ based on $\nabla_{\bm{V}}\mathcal{J}$, until convergence is achieved (Lines 9-22). 
In AdaGrad-based updating steps (Lines 12-13 and 18-19), $\odot$ is the Hadamard product operator, and $\circ a$ denotes the element wise exponentiation, with the exponent being $a$. 

\begin{algorithm}[h]
\caption{Optimization Algorithm of MDMF}
\label{alg:MDMF}
\SetKwData{Left}{left}\SetKwData{This}{this}\SetKwData{Up}{up}
\SetKwInOut{Input}{input} \SetKwInOut{Output}{output}
\Input{$\bm{Y}$, $D$, $T$, $\{\bm{S}^{d,h}\}_{h=1}^{m_d}$, $\{\bm{S}^{t,h}\}_{h=1}^{m_t}$, $\theta$, $k$, $opti$={\textit{GD} or \textit{AdaGrad}}}
\Output{$\bm{U}$, $\bm{V}$} 
Initialize $\bm{U}$ and $\bm{V}$ randomly\;
\For {$i\leftarrow 1$ \KwTo $m_d$}{ 
  Compute $\hat{\bm{S}}^{d,i}$, $\bm{L}^d_i$ and $w^d_i$\;
}
\For {$j\leftarrow 1$ \KwTo $m_t$}{ 
  Compute $\hat{\bm{S}}^{t,j}$, $\bm{L}^t_j$ and $w^t_j$\;
}
Compute $\bar{\bm{M}}$ \;
\If{opti=\text{AdaGrad}}{
$\bm{F},\bm{H} \leftarrow \bm{0}, \bm{0}$ \tcc*[r]{accumulated gradients used by AdaGrad}
}
\Repeat{convergence}{
    Compute $\nabla_{\bm{U}}\mathcal{J}$ according to Eq.  \eqref{eq:deriv_U} \;
    \eIf{opti=\text{AdaGrad}}{
        $\bm{F} \leftarrow \bm{F} + \left(\nabla_{\bm{U}}\mathcal{J} \right)^{\circ2}$ \;
        $\bm{U} \leftarrow \bm{U} - \theta \nabla_{\bm{U}}\mathcal{J} \odot \bm{F}^{\circ-\frac{1}{2}}$ \;     
        }{ $\bm{U} \leftarrow \bm{U} - \theta \nabla_{\bm{U}}\mathcal{J}$ \;}
    Compute $\nabla_{\bm{V}}\mathcal{J}$ according to Eq. \eqref{eq:deriv_V} \;
    \eIf{opti=\text{AdaGrad}}{
        $\bm{H} \leftarrow \bm{H} + \left(\nabla_{\bm{V}}\mathcal{J} \right)^{\circ2}$ \;
        $\bm{V} \leftarrow \bm{V} - \theta \nabla_{\bm{V}}\mathcal{J} \odot \bm{H}^{\circ-\frac{1}{2}}$ \; 
    }
    { $\bm{V} \leftarrow \bm{V} - \theta \nabla_{\bm{V}}\mathcal{J}$ \;}
}
\end{algorithm}

\subsection{AUPR and AUC losses optimization}

To optimize the AUPR loss in MDMFAPUR, we first obtain the gradients of $\mathcal{L}_{AP}$ w.r.t. $\bm{U}$ and $\bm{V}$:
\begin{align}
\nabla_{\bm{U}} \mathcal{L}_{AP} = & \sum_{h=1}^{n_b} \frac{\psi_h}{ (\widetilde{\Psi}_h)^2} \left( \Psi_h \sum_{i=1}^h \widetilde{\bm{Z}}^{i} \bm{V} - \widetilde{\Psi}_h \sum_{i=1}^h \bm{Z}^{i}\bm{V} \right)  - \frac{\Psi_h}{ \widetilde{\Psi}_h} \bm{Z}^{h}\bm{V}
\end{align}
\begin{align} 
\nabla_{\bm{V}} \mathcal{L}_{AP} = & \sum_{h=1}^{n_b} \frac{\psi_h}{ (\widetilde{\Psi}_h)^2} \left( \Psi_h \sum_{i=1}^h \widetilde{\bm{Z}}^{{i}^{\top}}  \bm{U} - \widetilde{\Psi}_h \sum_{i=1}^h {\bm{Z}^{i}}^{\top} \bm{U} \right) - \frac{\Psi_h}{ \widetilde{\Psi}_h} {\bm{Z}^{h}}^{\top} \bm{U} ,
\end{align}
where $\widetilde{\psi}_h = \psi(\delta(\hat{\bm{Y}},h))$, $\psi_h = \psi(\delta(\hat{\bm{Y}},h)\odot \bm{Y})$,  $\Psi_h= \sum_{i=1}^h \psi_i$, $\widetilde{\Psi}_h= \sum_{i=1}^h \widetilde{\psi}_i$, $\widetilde{\bm{Z}}^{h} = \nabla_{\bm{\hat{Y}}} \delta(\bm{\hat{Y}},h) \odot \hat{\bm{Y}} \odot (\bm{1}-\hat{\bm{Y}})$, $\bm{Z}^h =\bm{Y} \odot \widetilde{\bm{Z}}^{h}$. 
Then, $\nabla_{\bm{U}}\mathcal{L}$ and  $\nabla_{\bm{V}}\mathcal{L}$ are replaced by $\nabla_{\bm{U}} \mathcal{L}_{AP}$ and $\nabla_{\bm{U}} \mathcal{L}_{AP}$ in the computation of $\nabla_{\bm{U}}\mathcal{J}$ and $\nabla_{\bm{V}}\mathcal{J}$ (Line 10 and 16), respectively.
Furthermore, the simple gradient descent is utilized by MDMFAUPR due to its computational efficiency in Algorithm \ref{alg:MDMF}.

Concerning the optimization of MDMFAUC, the gradients of $\mathcal{L}_{AUC}$ w.r.t. $\bm{U}$ and $\bm{V}$ are:
\begin{equation}
\begin{aligned}
    \nabla_{\bm{U}} \mathcal{L}_{AUC} & = \left[\left(\nabla_{\bm{U}_1} \mathcal{L}_{AUC}\right)^\top,  \dotsc, \left(\nabla_{\bm{U}_{n_d}} \mathcal{L}_{AUC}\right)^\top  \right]^\top   \\ 
    \nabla_{\bm{U}_i} \mathcal{L}_{AUC} & \approx  \sum_{(ij,hl) \in P'}  \phi'( \zeta_{ijhl} ) \bm{V}_{j}  -\sum_{(hl,ij)\in P'}  \phi'(\zeta_{hlij}) \bm{V}_{j}
\end{aligned}
\label{eq:Lauc_U} 
\end{equation}
\begin{equation}
\begin{aligned}
    \nabla_{\bm{V}} \mathcal{L}_{AUC} &= \left[\left(\nabla_{\bm{V}_1} \mathcal{L}_{AUC}\right)^\top,  \dotsc, \left(\nabla_{\bm{V}_{n_t}} \mathcal{L}_{AUC}\right)^\top  \right]^\top  \\ 
    \nabla_{\bm{V}_{j}} \mathcal{L}_{AUC} & \approx \sum_{(ij,hl) \in P'}  \phi'( \zeta_{ijhl} )\bm{U}_i -\sum_{(hl,ij)\in P'}  \phi'(\zeta_{hlij})\bm{U}_i,
\end{aligned}
\label{eq:Lauc_V} 
\end{equation}
where $P'$ is a $n_dn_t$ sized set sampled from $P$, which accelerates the computation of $\nabla_{\bm{U}} \mathcal{L}_{AUC}$ and $\nabla_{\bm{V}} \mathcal{L}_{AUC}$ to achieve linear complexity w.r.t. $n_dn_t$. Similar to DWMFAUPR, $\nabla_{\bm{U}} \mathcal{L}_{AUC}$ ($\nabla_{\bm{U}} \mathcal{L}_{AUC}$) substitutes $\nabla_{\bm{U}}\mathcal{L}$ ($\nabla_{\bm{V}}\mathcal{L}$) for calculating the gradient of objective function in Line 10 (Line 16) of Algorithm \ref{alg:MDMF}.
For MDMFAUC, we employ the AdaGrad algorithm that adaptively changes the learning rate based on the accumulation of previous gradients to diminish the influence of sampling on computing gradients of $\mathcal{L}_{AUC}$ in Algorithm \ref{alg:MDMF}.

\subsection{Complexity analysis}
The complexity of MDMF is dominated by updating $\bm{U}$ and $\bm{V}$, particularly, the computation of gradients, e.g. $\nabla_{\bm{U}} \mathcal{L}$ and $\nabla_{\bm{V}} \mathcal{L}$. In each iteration, the cost of obtaining $\nabla_{\bm{U}}\mathcal{J}$ and $\nabla_{\bm{V}}\mathcal{J}$ is $O(\nabla_{\bm{U}}\mathcal{L})+O(r(m_dn_d^2+n_dn_t))$ and 
$O(\nabla_{\bm{V}}\mathcal{L})+O(r(m_tn_t^2+n_dn_t))$ respectively, where $O(\nabla_{\bm{U}}\mathcal{L})$ ($O(\nabla_{\bm{V}}\mathcal{L})$) is the complexity of calculating the gradient of the loss function.

Considering $O(\nabla_{\bm{U}}\mathcal{L}_{AP})=O(\nabla_{\bm{V}}\mathcal{L}_{AP})=O(rn_bn_dn_t)$ for the AUPR loss, the overall computational cost of MDMFAUPR is $O(rn_bn_dn_t)$, as $n_bn_dn_t>m_dn_d^2$ and $n_bn_dn_t>m_dn_d^2$. The complexity of MDMFAUC, which utilizes $\mathcal{L}_{AUC}$ with $O(\nabla_{\bm{U}}\mathcal{L}_{AUC})=O(\nabla_{\bm{V}}\mathcal{L}_{AUC})=O(rn_dn_t)$, is $O(r*\max\{m_dn_d^2,m_tn_t^2\})$. Lastly, MDMF2A, which integrates MDMFAUPR and MDMFAUC, requires a computational cost of $O(r(n_bn_dn_t+\max\{m_dn_d^2,m_tn_t^2\}))$.

\section{Baseline Models}
We introduce the baselines methods used in our experiments.
\begin{itemize}
    \item Five DTI prediction methods based on conventional machine learning techniques.
    \begin{itemize}
        \item \textbf{WkNNIR}~\cite{Liu2021Drug-targetRecovery} utilizes the recovered interactions inferred from neighborhood information, i.e., interactions of drugs having similar chemical structures and targets possessing similar protein sequences, to predict new DTIs.
        \item \textbf{NRLMF}~\cite{Liu2016NeighborhoodPrediction} trains a logistic matrix factorization model with neighborhood regularization on chemical and sequence based similarities, and assigns a higher weight to the minority interacting data.
        \item \textbf{MSCMF}~\cite{Zheng2013CollaborativeInteractions} learns two low-rank features that align to linear combinations of several drug and target similarity matrices derived from diverse data sources, respectively. 
        \item \textbf{GRGMF}~\cite{Zhang2020ANetworks} is a generalized matrix factorization model that employs an adaptive learning algorithm to refine the neighbourhood information and leverages the graph regularization to preserve the local invariance of learned latent drug and target features.
        \item \textbf{MF2A}~\cite{liu2021optimizing} linearly integrates multiple similarities with a weighting scheme that boosts the similarity with the higher local interaction consistency (LIC), and assembles two base models that optimize AUPR and AUC respectively.
        \item \textbf{DRLSM}~\cite{Ding2020IdentificationFusion} employs multiple kernel learning based on the Hilbert–Schmidt Independence Criterion to integrate multiple drug and target similarities into two spaces (drug and target), upon which a dual Laplacian regularized least squares classifier is trained to make predictions.
    \end{itemize}
    \item Two network-based DTI prediction approaches using random walk embedding methods.
     \begin{itemize}
     \item \textbf{DTINet}~\cite{Luo2017AInformation} is a network based model. It makes use of multiple heterogeneous data sources to generate drug and target embeddings by applying random walks with restart (RWR) and diffusion component analysis. Then, it infers new interactions by projecting drug and target embeddings in the same space by measuring the geometric proximity.
    \item \textbf{NEDTP}~\cite{An2021AInteractions} is a network based model. It generates drug and target embeddings using a Node2Vec model with random walks sampled from diverse drug and target related sub-networks. Interaction prediction is accomplished by training a Gradient Boosting Decision Tree classifier using the binary dataset constructed upon the learned embeddings.
    \end{itemize}
    \item Four deep learning DTI prediction models.
    \begin{itemize}
        \item \textbf{NeoDTI}~\cite{Wan2019NeoDTI:Interactions} implements a graph convolution architecture to leverage information from multiple heterogeneous sources and conducts the network reconstruction to discover new DTIs. 
        \item \textbf{DTIP}~\cite{Xuan2021IntegratingPrediction} uses a feature-level attention mechanism to extract fine-grained information for drug, target, and disease nodes. It also conducts RWR in the heterogeneous network to acquire node sequences, which are processed by a Bi-GRU based model with attention mechanism to generate the drug and target embeddings. 
        \item \textbf{DCFME}~\cite{chen2022DCFME} leverages CNN and MLP to learn drug and target features that capture both global/local and deep representations of drug-target couplings, and employs the focal loss to emphasize the hard examples during the training procedure.  
        \item \textbf{SupDTI}~\cite{chen2022SupDTI}, a deep learning DTI prediction framework enhanced by the self-supervised learning strategy, incorporates a contrastive learning and a generative learning modules to improve the global and local node-level agreement of learned embeddings in the heterogeneous DTI network. 
    \end{itemize}
\end{itemize}

In addition, the average of the corresponding multiple similarities are used as input for WkNNIR, NRLMF, and GRGMF, since they can only handle a single type of drug and target similarity. 
NetMF~\cite{qiu2018network}, applicable to single-layer networks only, works on the holistic DeepWalk matrix $\bar{M}$ that incorporates information from multiple hyper-layers. Multi2Vec~\cite{Teng2021AEmbedding}, designed for multiplex networks, learns embeddings on the DTI network composed of multiple hyper-layers, and averages the node vector representations across all hyper-layers as the final output embeddings.
For both network embedding models, after the embedding learning process, we construct a binary training dataset, where each training drug-target pair $(d_i,t_j)$ is transformed to an instance, whose feature vector is the concatenation of the embeddings of $d_i$ and $t_j$ and the label is $Y_{ij}$, and then employ it to train a random forest (RF) classifier with 100 trees using the Scikit-learn library~\cite{Pedregosa2011Scikit-learn:Python}. A binary test dataset is built upon test pairs in a similar way, and the RF is used to predict the test set.
Network embedding methods follow the transductive learning scheme, as test drugs (targets) should be available and utilized in the embedding learning process.

\section{Parameter Setting}
The parameter settings of MDMF2A for each dataset and prediction setting are listed in Supplementary Table \ref{tab:param}.

\begin{table}[h]
\centering
\caption{The best hyperparameter settings of MDMF2A.}
\label{tab:param}
\begin{tabular}{cc|c|ccccc|cccc}
\toprule
\multirow{2}{*}{Setting} & \multirow{2}{*}{DataSet} & \multirow{2}{*}{$\beta$} & \multicolumn{5}{c|}{MDMFAUPR} & \multicolumn{4}{c}{MWMFAUC} \\
 &  &  & $r$ & $\lambda_d$ & $\lambda_t$ & $\lambda_r$ & $n_b$ & $r$ & $\lambda_d$ & $\lambda_t$ & $\lambda_r$ \\ \hline
\multirow{5}{*}{S1} & NR & 0.91 & 100 & $2^{-4}$ & $2^{-4}$ & $2^{-6}$ & 31 & 50 & $2^{0}$ & $2^{0}$ & $2^{-4}$ \\
 & GPCR & 0.57 & 100 & $2^{0}$ & $2^{-4}$ & $2^{-4}$ & 16 & 100 & $2^{0}$ & $2^{-4}$ & $2^{-4}$ \\
 & IC & 0.8 & 100 & $2^{-4}$ & $2^{0}$ & $2^{-4}$ & 11 & 50 & $2^{2}$ & $2^{0}$ & $2^{-4}$ \\
 & E & 0.56 & 100 & $2^{0}$ & $2^{0}$ & $2^{-4}$ & 21 & 100 & $2^{2}$ & $2^{2}$ & $2^{-4}$ \\
 & Luo & 0.88 & 100 & $2^{-4}$ & $2^{-4}$ & $2^{-4}$ & 21 & 100 & $2^{2}$ & $2^{2}$ & $2^{0}$ \\ \hline
\multirow{5}{*}{S2} & NR & 0.9 & 50 & $2^{-6}$ & $2^{-6}$ & $2^{-6}$ & 16 & 50 & $2^{2}$ & $2^{-4}$ & $2^{-4}$ \\
 & GPCR & 0.63 & 100 & $2^{0}$ & $2^{-6}$ & $2^{-6}$ & 11 & 100 & $2^{0}$ & $2^{0}$ & $2^{-4}$ \\
 & IC & 0.52 & 100 & $2^{-4}$ & $2^{-4}$ & $2^{-4}$ & 11 & 50 & $2^{2}$ & $2^{0}$ & $2^{-4}$ \\
 & E & 0.72 & 100 & $2^{-4}$ & $2^{-4}$ & $2^{0}$ & 21 & 100 & $2^{2}$ & $2^{2}$ & $2^{2}$ \\
 & Luo & 0.89 & 100 & $2^{0}$ & $2^{-4}$ & $2^{-4}$ & 21 & 100 & $2^{2}$ & $2^{2}$ & $2^{-4}$ \\ \hline
\multirow{5}{*}{S3} & NR & 0.99 & 100 & $2^{0}$ & $2^{-6}$ & $2^{-4}$ & 21 & 100 & $2^{2}$ & $2^{0}$ & $2^{-4}$ \\
 & GPCR & 0.83 & 50 & $2^{0}$ & $2^{-4}$ & $2^{-4}$ & 26 & 100 & $2^{-4}$ & $2^{-4}$ & $2^{0}$ \\
 & IC & 0.58 & 100 & $2^{-4}$ & $2^{-4}$ & $2^{0}$ & 26 & 100 & $2^{0}$ & $2^{0}$ & $2^{0}$ \\
 & E & 0.77 & 100 & $2^{-6}$ & $2^{0}$ & $2^{0}$ & 21 & 100 & $2^{2}$ & $2^{2}$ & $2^{-4}$ \\
 & Luo & 0.8 & 100 & $2^{-4}$ & $2^{-4}$ & $2^{-4}$ & 21 & 100 & $2^{2}$ & $2^{2}$ & $2^{-4}$ \\ \hline
\multirow{5}{*}{S4} & NR & 0.97 & 50 & $2^{-4}$ & $2^{-4}$ & $2^{-6}$ & 31 & 100 & $2^{2}$ & $2^{-4}$ & $2^{0}$ \\
 & GPCR & 0.52 & 100 & $2^{0}$ & $2^{-6}$ & $2^{-6}$ & 16 & 100 & $2^{2}$ & $2^{-4}$ & $2^{0}$ \\
 & IC & 0.85 & 50 & $2^{-4}$ & $2^{-6}$ & $2^{-4}$ & 11 & 50 & $2^{0}$ & $2^{0}$ & $2^{0}$ \\
 & E & 0.79 & 100 & $2^{0}$ & $2^{0}$ & $2^{-4}$ & 21 & 100 & $2^{2}$ & $2^{2}$ & $2^{0}$ \\
 & Luo & 0.93 & 100 & $2^{-6}$ & $2^{-4}$ & $2^{-6}$ & 21 & 100 & $2^{2}$ & $2^{2}$ & $2^{0}$ \\ \bottomrule
\end{tabular}
\end{table}

\section{Ablation Study}

\begin{figure}[h]
\centering
\subfloat[AUPR]{\includegraphics[width=0.5\textwidth]{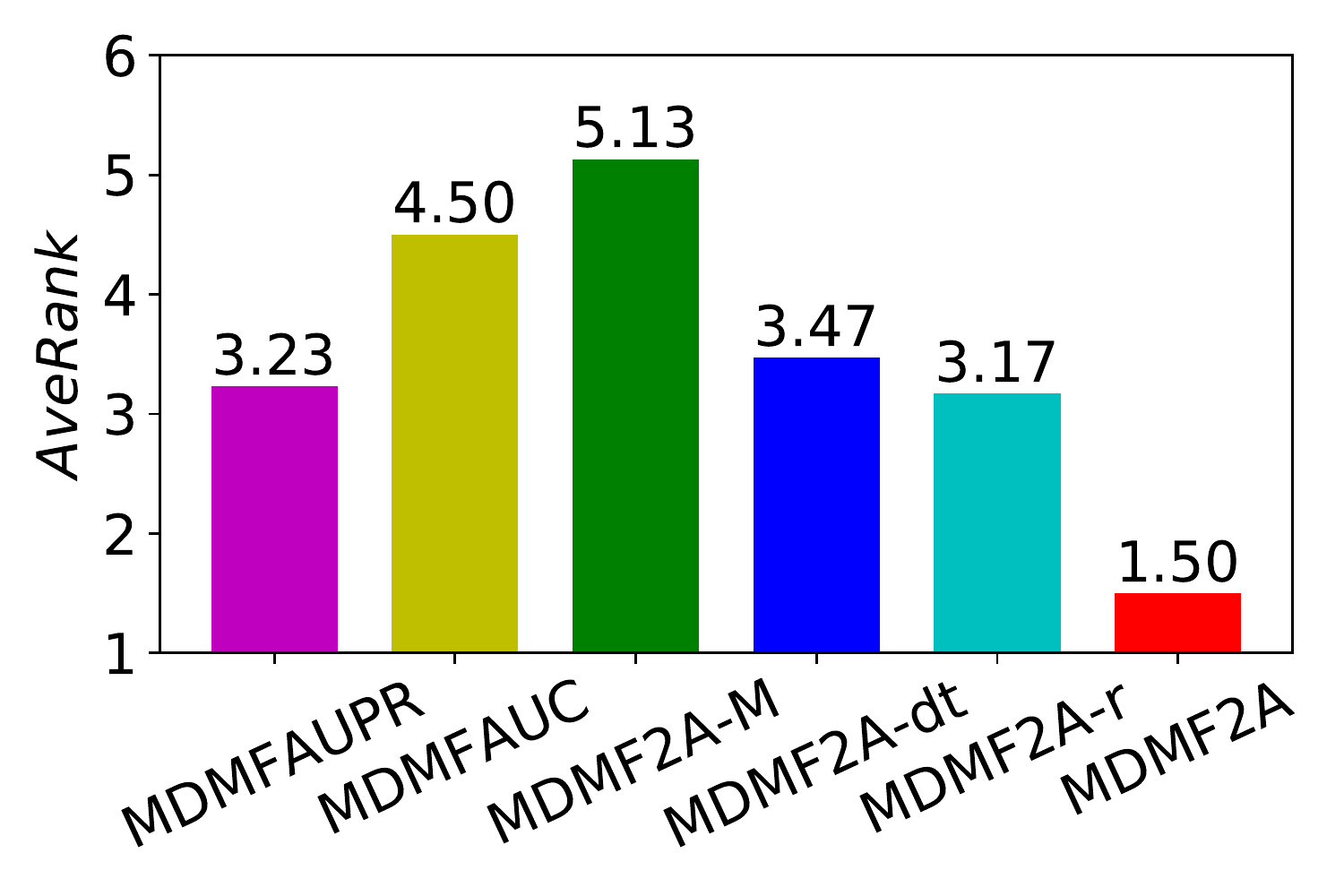}}
\subfloat[AUC]{\includegraphics[width=0.5\textwidth]{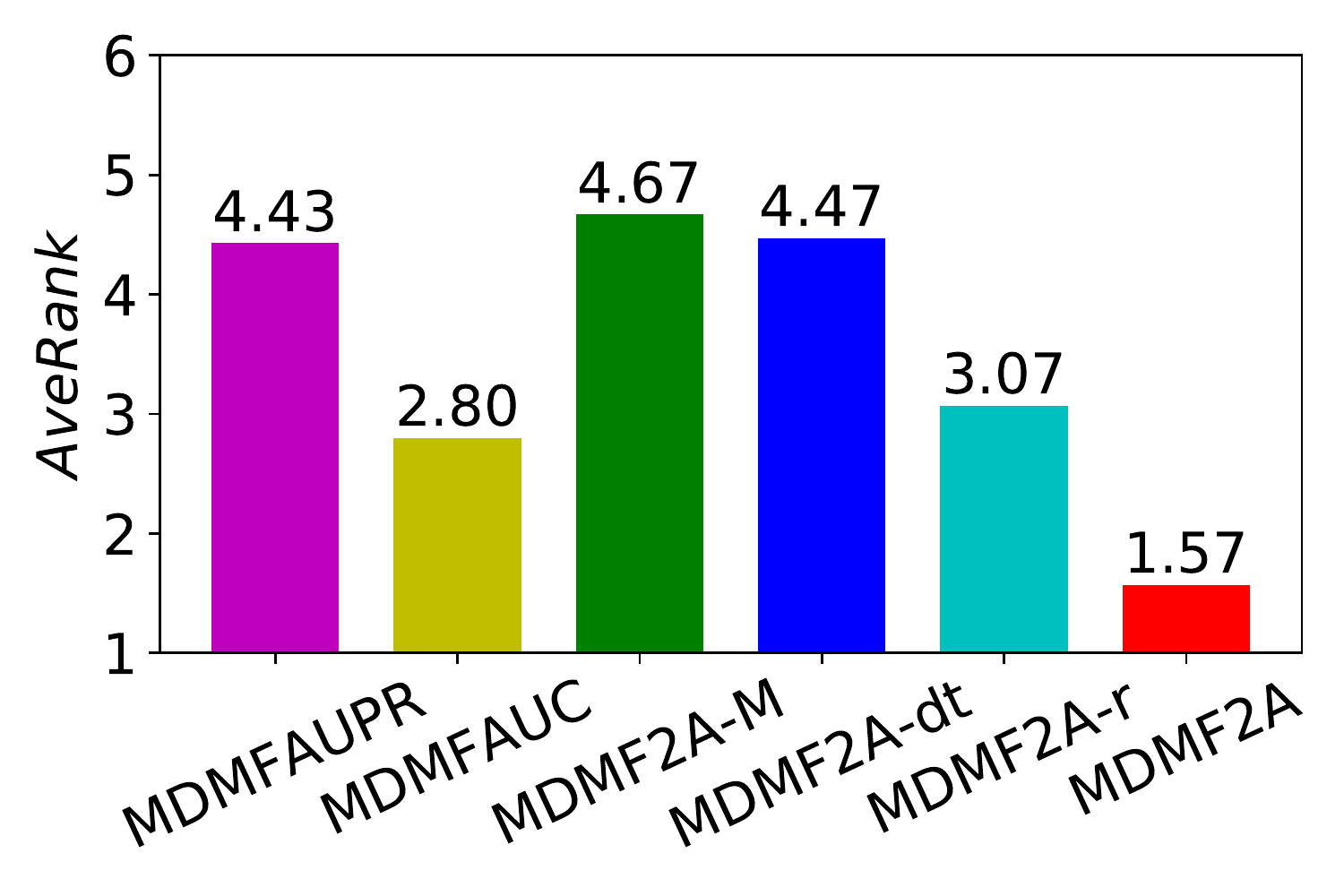}}
\caption{The average rank of MDMF2A and its degenerate variants over all five datasets and four settings. Lower rank indicates better performance.} 
\label{fig:Ablantion_AveRank}
\end{figure}

To demonstrate the effectiveness of the ensemble framework and the regularization terms employed in the proposed method, we conduct an ablation study on MDMF2A by considering its five degenerate variants: (i) MDMFAUPR by setting $\beta$ as 1; (ii) MDMFAUC by setting $\beta$ as 0; (iii) MDMF2A-M excludes DeepWalk matrix based regularization, i.e., $\lambda_M=0$; (iv) MDMF2A-dt does not consider layer-specific graph regularization i.e., $\lambda_{d}=\lambda_{t}=0$; (v) MDMF2A-r ignores Tikhonov regularization, i.e., $\lambda_r=0$. 
The average rank of MDMF2A and its five variants over all five datasets and four settings are shown in Figure~\ref{fig:Ablantion_AveRank}. 
Firstly, although MDMFAUPR and MDMFAUC perform well in terms of the metric they optimize, they become worse in the other metric. MDMF2A outperforms its two base models in terms of both AUC and AUPR, verifying the success of integrating two single metric aware models in boosting performance. Concerning the regularization terms, we observe that  removal of any term would trigger severe performance degradation, which confirms their importance in MDMF-based models. 
More specifically, DeepWalk matrix-based regularization, which captures high-order proximity from network topology, is the most crucial one. Layer-specific graph regularization that perverse distinctive information from each view comes next, and it contributes more to AUC. Although the damage caused by neglecting Tikhonov regularization is less than the other two, its function to avoid overfitting cannot be ignored.

\begin{figure}[h]
\centering
\subfloat[S1]{\includegraphics[width=0.43\textwidth]{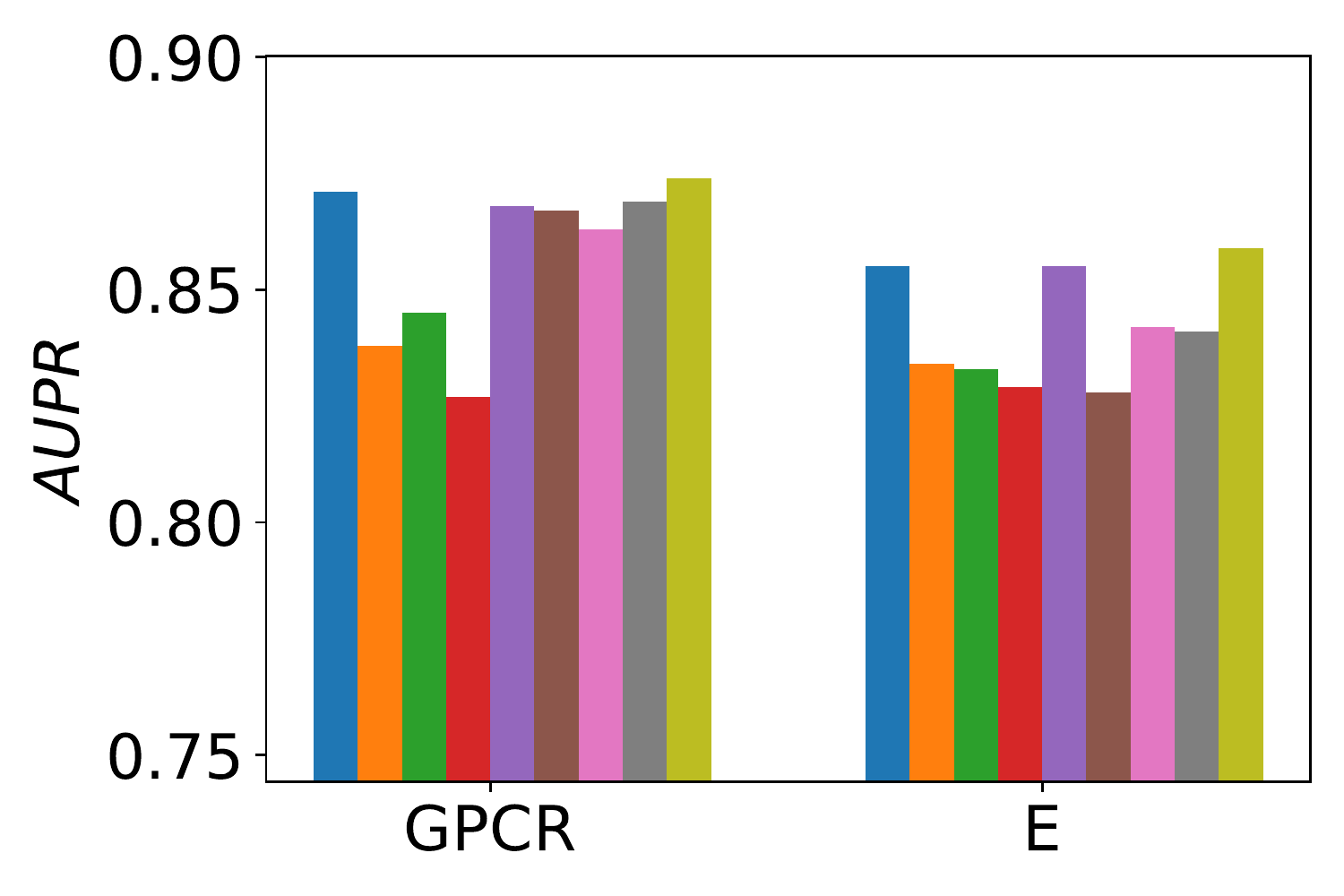}}
\hspace{1em}
\subfloat[S2]{\includegraphics[width=0.43\textwidth]{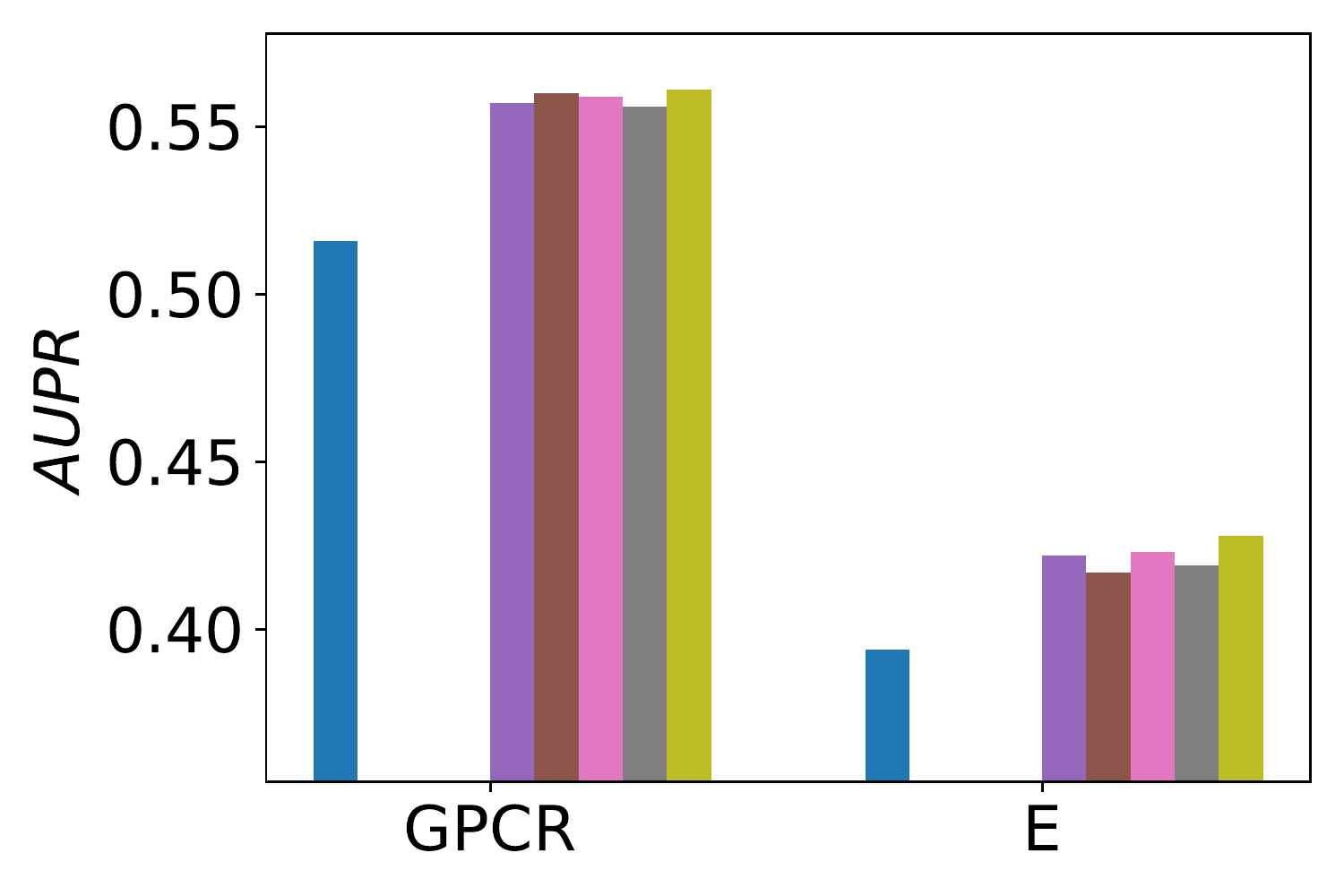}} \\
\subfloat[S3]{\includegraphics[width=0.43\textwidth]{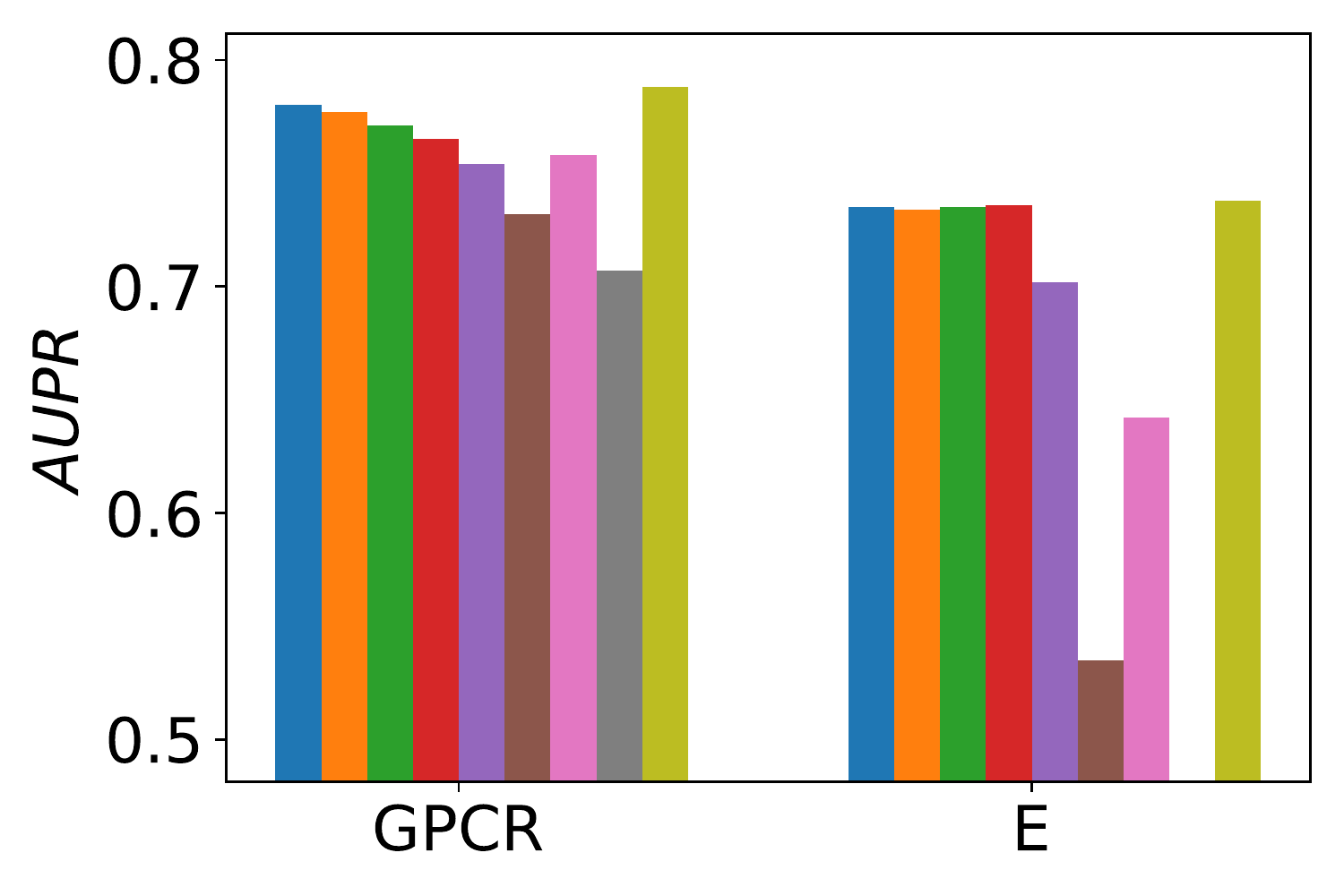}}
\hspace{1em}
\subfloat[S4]{\includegraphics[width=0.43\textwidth]{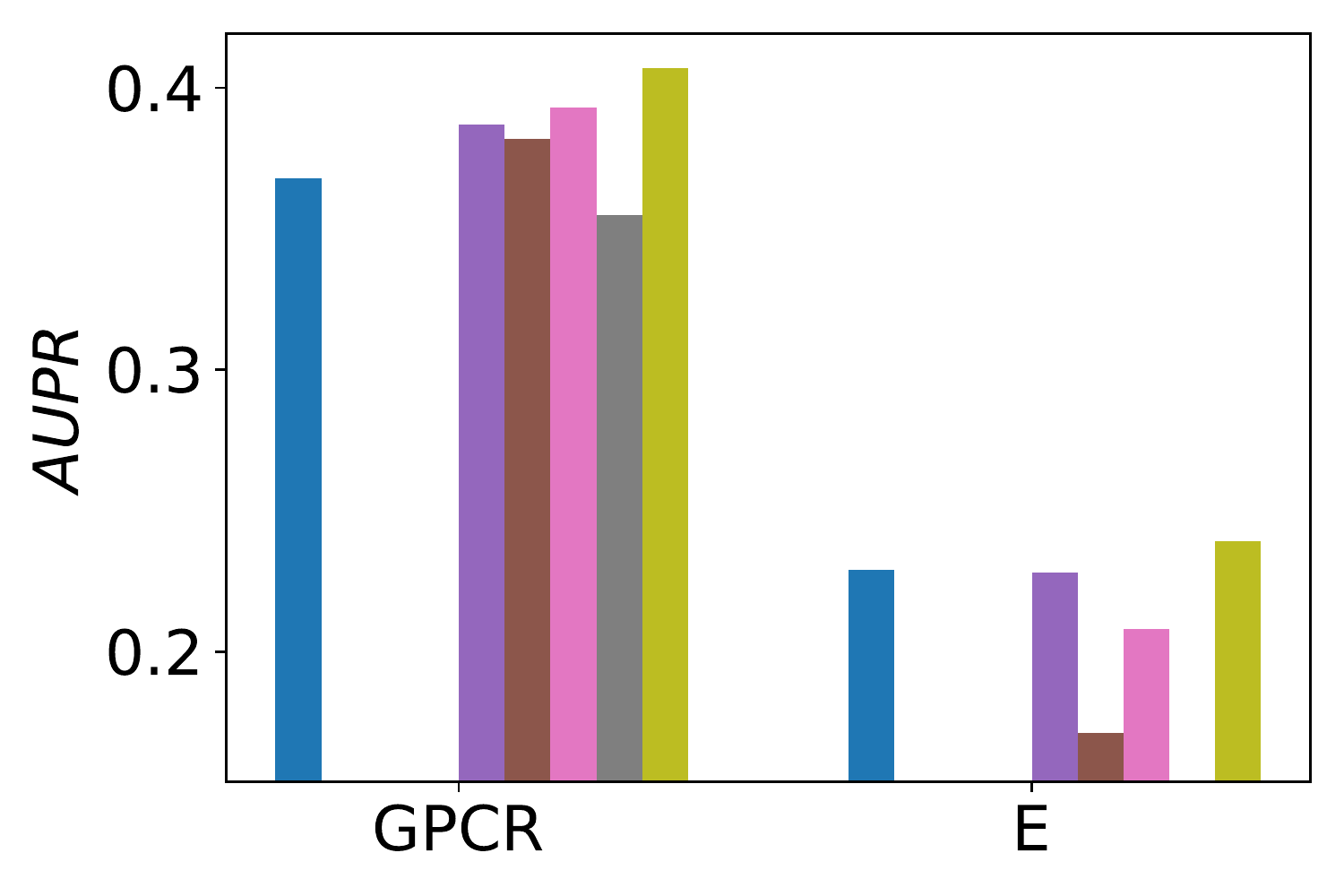}} \\
\subfloat{\includegraphics[width=0.89\textwidth]{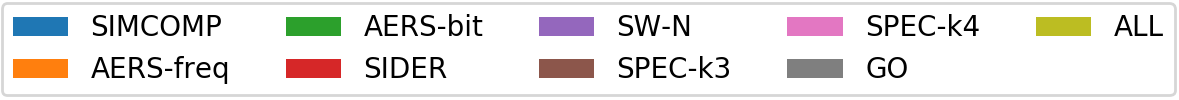}} \\
\caption{AUPR results of MDMF2A using various types of similarities on \textsl{GPRC} and \textsl{E} datasets, where sub-figures (a), (b), (c), and (d) show the results in prediction setting S1, S2, S3, and S4, respectively.} 
\label{Sfig:Single_Sim}
\end{figure}

We further compare MDMF2A using multiple similarity types and each type alone. Supplementary Figure \ref{Sfig:Single_Sim} shows AUPR results of MDMF2A using different type(s) of drug and target similarities. ALL denotes using all four types of drug similarities and four types of target similarities. SIMCOMP, AERS-freq, AERS-bit and SIDER denote using a single type of drug similarities and four types of target similarities. Likewise, SW-N, SPEC-k3, SPEC-k4 and GO indicate leveraging one type of target similarities and all kinds of drug similarities. 
For the three side effect based drug similarities, since their corresponding side effect databases do not record all drugs in updated Golden Standard datasets, the similarities between a drug without any known side effect and other drugs are defined as zeros. In other words, given a side effect based drug similarity matrix, elements in the rows corresponding to drugs not presented in AERS (SIDER) database are zeros.
The existence of zero similarity rows leads to the invalidity of latent feature inference for new drugs defined in Eq.(8), because $\bm{\bar{s}}^d_{x}=\bm{0}$ and $\bm{U}_x$ would always be a zero-valued vector. Therefore, AERS-freq, AERS-bit and SIDER drug similarities can not be individually applied to S2 and S4, where interactions of new drugs need to be estimated.
Likewise, GO target similarity of E dataset contains several zero rows, so it can not be used to predict interactions of new targets (S3 and S4). From Figure \ref{Sfig:Single_Sim}, we can see that integrating multiple similarities can capture more useful information than using single similarities, leading to better performance.

\section{Parameter Analysis}
\begin{figure}[h]
\centering
\subfloat[\textsl{NR} dataset]{\includegraphics[width=0.49\textwidth]{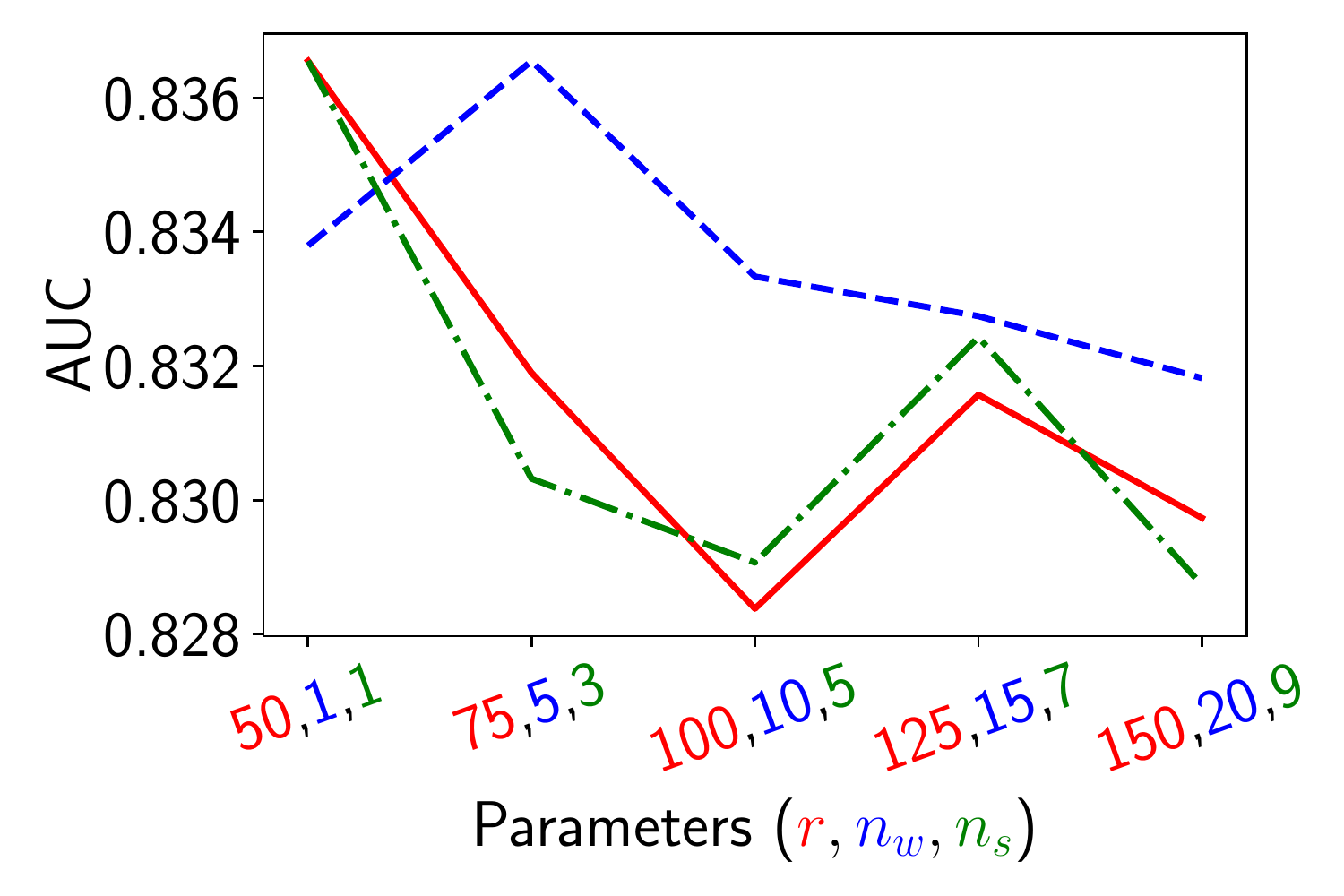}}
\subfloat[\textsl{E} dataset]{\includegraphics[width=0.49\textwidth]{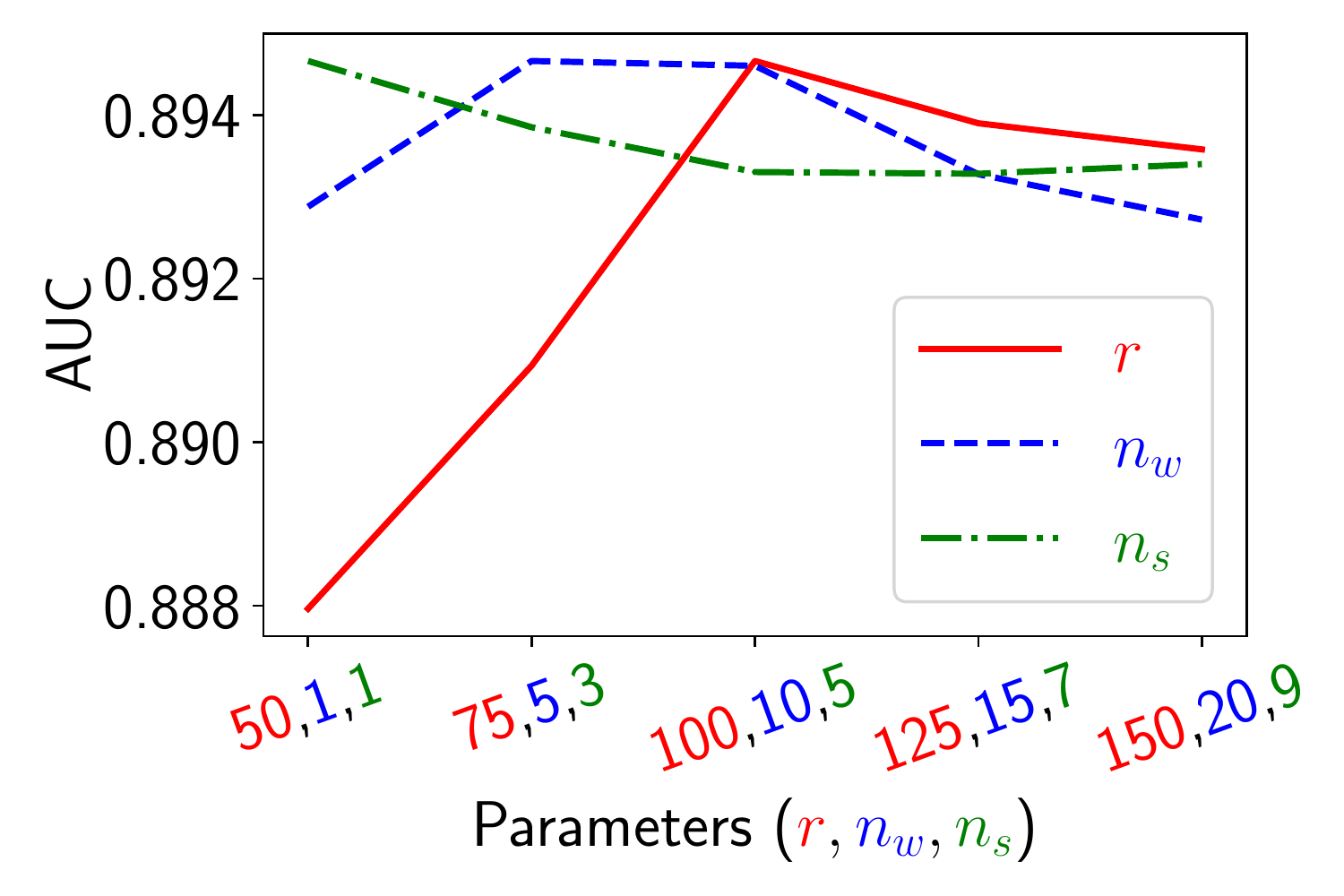}}
\caption{AUC results of MDMF2A under different parameter configurations on \textsl{NR} and \textsl{E} dataset in S2.} 
\label{fig:VariousParams}
\end{figure}

We further investigate the sensitivity of three important parameters: $r$, $n_w$, and $n_s$. Figure~\ref{fig:VariousParams} shows the AUC results of MDMF2A under different parameter configurations on one smaller (\textsl{NR}) and one larger (\textsl{E}) dataset in S2. 

Concerning the embedding dimension, $r$, lower values lead to better performance for the small-sized \textsl{NR} dataset. In the larger \textsl{E} dataset, the AUC results improve with higher values and plateaus when $r$ reaches 100. For $n_w$, its optimal value is around 5 for both datasets. A smaller window size cannot exploit sufficient proximity among nodes, while a larger one incorporates unrelated nodes in the context, leading to performance deterioration. MDMF2A achieves the best performance when $n_s=1$, and becomes worse when $n_s$ increases, because a larger number of negative samples leads to a more sparse DeepWalk matrix, which cannot capture the structural information adequately. The impact of sparsity caused by larger $n_s$ is more severe for the smaller NR dataset, whose corresponding DeepWalk matrix contains fewer non-zero values. Furthermore, the proposed model is more insensitive to DeepWalk related parameters, e.g., $n_w$ and $n_s$, than the embedding dimension on \textsl{E} dataset. This demonstrates the robustness of MDMF2A to the parameter variation for mining large-scale networks.

\clearpage
\bibliographystyle{unsrt}
\bibliography{BL_copy,ref}